\documentclass[review]{elsarticle}

\usepackage{lineno,hyperref}
\modulolinenumbers[5]
\usepackage[margin=1in]{geometry}

\usepackage{graphicx}
\usepackage{listings}
\usepackage{makecell}
\usepackage[utf8]{inputenc}
\usepackage[T1]{fontenc}
\usepackage{amsmath}

\usepackage{booktabs}
\usepackage{algorithm}

\usepackage{algpseudocode}

\usepackage{tcolorbox}
\usepackage{listings}
\usepackage{caption}

\usepackage{sectsty}
\sectionfont{\fontsize{10}{10}\selectfont}

\usepackage{listings}
\usepackage{xcolor}

\usepackage[numbers,sort&compress]{natbib}
\biboptions{numbers,sort&compress}

\lstdefinestyle{promptstyle}{
    basicstyle=\ttfamily\small,
    breaklines=true,
    frame=single,
    columns=fullflexible,
    keepspaces=true,
    showstringspaces=false
}

\usepackage{amssymb}
\usepackage{lscape}
\usepackage{tabularx}

\begin{document}

\begin{frontmatter}



\title{Policy-Guided Threat Hunting: An LLM enabled Framework with Splunk SOC Triage}



\author[1]{Rishikesh Sahay}
\ead{rsaha@uis.edu}

\author[1]{Bell Eapen}
\ead{bpunn@uis.edu}

\author[2]{Weizhi Meng}
\ead{w.meng3@lancaster.ac.uk}

\author[1]{Md Rasel Al Mamun}
\ead{mmamu@uis.edu}

\author[3]{Nikhil Kumar Dora}
\ead{2481140@kiit.ac.in}

\author[1]{Manjusha Sumasadan}
\ead{msuma5@uis.edu}

\author[3]{Sumit Kumar Tetarave}
\ead{sumitkumar.fca@kiit.ac.in}


\author[5]{Elyson De La Cruz}
\ead{elyson.delacruz@ucumberlands.edu}

\address[1]{Department of Management Information Systems, University of Illinois, Springfield, USA}

\address[2]{School of Computing and Communications, Lancaster University, United Kingdom}

\address[3]{School of Computer Applications,
Kalinga Institute of Industrial Technology, India}



\address[5]{School of Information Technology and Artificial Intelligence, University of Cumberlands, USA}

\begin{abstract}

With frequently evolving Advanced Persistent Threats (APTs) in cyberspace, traditional security solutions approaches have become inadequate for threat hunting for organizations. Moreover, SOC (Security Operation Centers) analysts are often overwhelmed and struggle to analyze the huge volume of logs received from diverse devices in organizations. 
To address these challenges, we propose an automated and dynamic threat hunting framework for monitoring evolving threats, adapting to changing network conditions, and performing risk-based prioritization for the mitigation of suspicious and malicious traffic.
By integrating Agentic AI with Splunk, an established SIEM platform, we developed a unique threat hunting framework.
The framework systematically and seamlessly integrates different threat hunting modules together, ranging from traffic ingestion to anomaly assessment using a reconstruction-based autoencoder, deep reinforcement learning (DRL) with two layers for initial triage, and a large language model (LLM) for contextual analysis. 
We evaluated the framework against a publicly available benchmark dataset, as well as against a simulated dataset. 
The experimental results show that the framework can effectively adapt to different SOC objectives autonomously and identify suspicious and malicious traffic.
The framework enhances operational effectiveness by supporting SOC analysts in their decision-making to block, allow, or monitor network traffic.
This study thus enhances cybersecurity and threat hunting literature by presenting the novel threat hunting framework for security decision- making, as well as promoting cumulative research efforts to develop more effective frameworks to battle continuously evolving cyber threats.

\end{abstract}



\begin{keyword}

Threat hunting, Splunk, Security Operation Center, LLM, Agentic AI, Deep Reinforcement Learning, Autoencoder


\end{keyword}

\end{frontmatter}


\section{Introduction}
\label{sec:intro}

The frequently evolving threat landscape in cyberspace emphasizes the need for proactive and intelligent cyber threat hunting~\cite{ferrag2024revolutionizingcyberthreatdetection}.
According to Kaspersky (2024), there is an increase of 74\% in advanced persistent threats (APTs) in 2024 compared to 2023~\cite{kaspersky}.
Advanced Persistent Threats are cyber threats that use sophisticated tools and resources to exploit vulnerabilities in organizations. According to the Fortinet threat report 2025, the attempts to exploit newly found vulnerabilities have increased, and cybercriminals are using artificial intelligence (AI) for phishing, impersonation, and evasion tactics, resulting in the increase of reconnaissance activity by 16.7\% each year~\cite{fortinet}.
Traditional security approaches have become inadequate with the rise of these advanced persistent threats because traditional endpoint detection and response tools rely on a known attack signature or a clear anomalous pattern~\cite{llm6g,increasingsophistication}. Contemporary anomaly-based detection solutions are reactive, fail to address continuously evolving threat landscapes, and thereby require proactive threat hunting approaches~\cite{minmaxad,deepunsupervisedanomaly,adsg}. 
Thus, this study offers a novel, proactive threat hunting framework with improved effectiveness to assist security analysts in their decision-making to block, allow, or monitor network traffic.

Mitigation of advanced persistent threats requires continuous monitoring of security logs. 
Thus, the security analysts in the security operation center (SOC) continuously analyze a large volume of traffic logs to effectively pinpoint potential threats~\cite{log_management}.
However, current threat hunting literature identifies that it is challenging for security analysts to analyze huge volume and complex data types of logs~\cite{threathuntingchallenges}.
Although the Security Information and Event Management (SIEM) tools such as Splunk offer centralized log aggregation, correlation, and real-time monitoring features, they rely on predefined rules and existing signatures, which limit their effectiveness against new or context-driven threats~\cite{splunk_stack,increasingsophistication}. In addition, it is imperative for the SOC analysts to perform risk-based prioritization for mitigation. 
Another critical issue for the current threat hunting process is the shortage of security analysts and the lack of automation in repetitive security workflows that causes a burden on SOC analysts~\cite{threathuntingchallenges}.
Thus, there is a need for proactive cybersecurity solutions with automatic security workflows in threat hunting that can monitor evolving threats and adapt to changing network conditions, perform risk-based prioritization for mitigation of suspicious and malicious traffic. 
Thus, integrating Agentic AI with Splunk, an established SIEM platform, this research proposes an automated threat hunting framework capable of identifying and mitigating evolving threats with high accuracy and validate the framework. 

The recent advancements in the Agentic AI system built on LLM (Large Language Model) present a significant opportunity to improve threat hunting operations~\cite{trismagenticaireview}.
The Agentic AI system leverages advancements in reinforcement learning, goal-oriented architectures, and adaptive control mechanisms~\cite{agentic-ai-survey}.
The agentic system comprises collaborative agents with specialized roles such as planning, analysis, and execution facilitated by LLM along with tool use.
In the Agentic AI architecture, LLM is the main decision making controller, also referred to as the brain of the system~\cite{agentic-ai-survey}. 
LLMs have shown considerable capabilities to identify complex patterns, analyze unstructured data, and provide contextual insights that are highly relevant to security operations~\cite{ferrag2024revolutionizingcyberthreatdetection,llmfor_cyber}.
Agentic AI is goal oriented, with adaptable features that enable it to complete multi-layered tasks without instructions each time~\cite{agentic-ai-survey}. 

By integrating the Agentic AI system with established SIEM platforms such as Splunk within the threat hunting workflow, we can automate log analysis, detect subtle indicators of compromise (IoCs), and reduce the burden on SOC analysts~\cite{gen_ai,questionsfordecisionsupport}. 
Recent works~\cite{transformingcybersecurity,simbianwhitepaper,autonoumousaiagent} highlight AI agents as assisting the SOC analyst by correlating logs across diverse data sources, preserving investigative context through memory, and continuously updating hypotheses over time.
It has the capability to automate contextual analysis, generate detection rules, assist in formulating sophisticated queries for security information and event management systems, and even support in the development of incident response playbook~\cite{automate_detection}. 
This integration of Agentic AI with the SIEM platform can facilitate real-time monitoring and alerting, allowing faster response times and decision making to security incidents~\cite{electronics13234718}.

Automating the threat hunting workflow allows SOC analysts to focus on strategic and innovative aspects of threat hunting~\cite{turninghuntedhunterthrea}. 
This proactive approach not only strengthens an organization's security posture, but also optimizes the utilization of limited security expertise.
However, in the SOC environment, human oversight is very important for safe autonomy and crucial decision-making. 
Under a fast changing environment and incomplete information, agents may struggle to generalize, so human-in-the-loop is necessary to validate inferred threats and ambiguous findings.
Therefore, the challenge is to design an Agentic threat hunting framework with coordinated workflow that keeps SOC analyst driven flexibility along with explainable decision making and providing broader automation without exceeding acceptable risk. 
After a comprehensive analysis of the available threat hunting processes, we have identified several requirements to enhance them, and even a new design: (1) minimize false alarms; (2) automation in key security workflows; (3) prioritization of traffic flows for further investigation by SOC analysts and LLM; (4) LLM assisted contextual analysis of traffic flows and generation of queries to filter logs on SIEM tools such as Splunk; (5) Automated development of incident response playbook; (6) SOC analyst involved in final decision making.

To meet all the above-mentioned requirements, we propose a framework for automated and dynamic threat hunting leveraging the capabilities of Agentic AI to address the changing threat landscape. 
The framework is intended to systematically and seamlessly integrate different threat hunting modules together, ranging from log ingestion to anomaly assessment using reconstruction-based autoencoder, Deep Reinforcement Learning with two layers, and LLM triage.
In particular after log collection, autoencoder based anomaly detection is trained on a part of initial benign traffic and assigns confidence score to all the traffic instances based on the learned normal network features.
In the framework, the Deep Reinforcement Learning (DRL) module is trained on the traffic of fixed length time window for decision making.
After DRL decision, traffic flows are prioritized considering DRL decision and autoencoder anomaly score for LLM analysis using ChatGPT.
Only flows with a high priority score are forwarded to LLM for contextual analysis to avoid unnecessary computational overload and hallucination.
Based on contextual insights from LLM, further validation is also performed on Splunk to identify malicious and suspicious activities related to flows with a high priority score.
This workflow significantly improves alert triage and reduces the burden on SOC analysts and helps them in informed decision-making.
Moreover, it is important to differentiate Agentic AI with the deep reinforcement‑learning (DRL) agent in our framework.
Agentic AI refers to a broader architectural paradigm in which an LLM orchestrates a collection of specialized, tool using agents that collaborate to achieve high‑level goals.
Within this architecture, the DRL agents provide outputs that the LLM‑based agents incorporate into planning and decision‑making.

The rest of the paper is organized as follows.
The related literature on agentic AI and LLM application in cybersecurity and the SIEM tool for threat hunting is described in Section~\ref{sec:related_works}.
A set of key observations and motivation are presented in Section~\ref{sec:key_observation}
Section~\ref{sec:genai} presents the Agentic AI driven threat-hunting framework, its different components.
Section~\ref{sec:methodology} describes the components and its functionality in detail.
The use case and threat model are described in Section~\ref{sec:use_case}.
Section~\ref{sec:experimentation_results} described the experimental results.
Section~\ref{sec:discussion} presents a discussion of the framework and its limitations, and finally Section~\ref{sec:conclusion_future_work} concludes the article with future work.

\section{Related Work}
\label{sec:related_works}

Agentic AI is reshaping threat hunting with adaptive and dynamic approach beyond the traditional alert driven detection system in the network.  
Specifically, in the network, these agents monitor traffic, identify malicious activities, trigger containment, or mitigate that to reduce the burden on security analysts.
Recently, a few works have explored the application of argentic AI in different cybersecurity domains, including autonomous incident response, cyber threat intelligence, autonomous monitoring, and adversarial cyber defense~\cite{surveyagenticaicybersecurity,transformingcybersecurity}.
We review three groups of literature: (1) Agentic AI for threat hunting and adaptive defense, (2) Agentic AI based network monitoring and anomaly detection, and (3) LLM and Generative AI for Security Operation Centers (SOC) tasks.


\begin{itemize}
    \item \textbf{Agentic AI for threat hunting and adaptive defense:} In~\cite{autonoumousaiagent}, the authors present an autonomous threat hunting using machine learning and Deep Reinforcement Learning methods for proactive threat detection.
    The framework performs traffic analysis using ML models such as RNN and CNN, then leverages Deep Reinforcement Learning for optimal threat hunting.
    The paper highlights that automation can reduce the burden on the SOC analyst, although the main aim of this work is to detect and respond based on DRL learning. 
    In~\cite{agenticai-selfhealing}, the authors use agentic AI for self-healing cyber systems with autonomous detection, mitigation and adaptation to evolving cyber threats.
    However, the main focus of this work is to improve threat detection using agentic AI and does not provide SIEM integration to support SOC analysts in decision making.
    In~\cite{unifiedframeworkai}, the authors proposed a conceptual framework for the integration of AI with human analysts for SOC environments.
    The framework focuses on organizational and workflow aspects rather than anomaly assessment, threat prioritization, and automated investigation.
    In~\cite{aictm, simbianwhitepaper}, the role of agentic AI in cybersecurity is presented along with the challenges of using these AI agents for cybersecurity, as threat environments evolve.
    As a result, they emphasize that human analysts must be in the loop for validating threats and analyzing complex findings.
    Unlike these works, our framework provides autoencoder based anomaly assessment, DRL-based traffic triage and prioritization, and LLM-assisted analysis while also integrating with SIEM tools such as Splunk for the deep investigation of malicious flows.
    
    \item \textbf{Agentic AI based Network Monitoring and Anomaly Detection:} NetMonAI~\cite{netmonai} proposed a scalable distributed networking monitoring framework combining packet-level and flow-level analysis.
    Each node in the architecture has an agent which captures the traffic, find anomalies and reason using LLMs.
    These agents work in an automated way and coordinate with a centralized controller which collects reports and provides human readable summaries.
    In~\cite{argos}, a time series based anomaly detection system is proposed for cloud infrastructure.
    In the framework, multiple agents collaborate to autonomously generate detection rules using LLM and improve detection accuracy.
    NetMonAI and ARGOS leverage agentic AI for scalable network monitoring and anomaly detection, but do not address the complete SOC workflow.
    These works focus on anomaly detection, but our framework provides the full cycle of anomaly assessment, initial triage, LLM assisted contextual analysis, and SIEM platform based investigation of malicious flows.
 \item \textbf{LLMs and Generative AI for SOC Operations:} Recently, the integration of LLMs into cybersecurity has gained significant attention due to their ability to automate manual tasks, improved contextual analysis, and decision making~\cite{canchatgpt}.
 In~\cite{sahay_genai} the authors proposed a framework for threat hunting using LLM and Splunk.
 The framework leverages LLM for the initial triage of security logs and then performs further investigation on Splunk.
 In~\cite{alarm_analysis_llm} the authors present the potential of LLMs to automate the analysis and triage of security alerts, highlighting how LLMs can provide contextual insight, reduce false positives, and assist SOC analysts in improving security operations.
 In~\cite{llmfornonsecurity}, the use of different LLMs is explored for threat hunting. 
 The main focus of the work is to determine whether LLM can generate effective queries for security tools such as Splunk and Elasticsearch to analyze logs.
 In~\cite{benchmarkingllm} the authors provide a framework called \texttt{LLM4Sec} to improve anomaly detection by fine-tuning LLMs. 
 The authors demonstrate extensive evaluation of five LLMs such as BERT, RoBERTa, DistilRoBERTa, GPT-2 and GPT-Neo, investigating their effectiveness in log analysis.  
 In~\cite{usingllmtoautomatectisoc} an AI system is presented, leveraging LLMs such as GPT-4 to automate the extraction of Indicators of Compromise (IoCs) and to construct relationship graphs from Cyber Threat Intelligence (CTI) reports, which minimizes the manual tasks of SOC analysts.
 Vinayak~\cite{vinayak} studied the application of LLMs in cyber threat hunting.
 These works mainly investigated the potential application of LLMs in analyzing security logs and finding IoCs for threat detection.
 Industry grade solutions such as Microsoft Security Copilot have enabled natural language interaction and contextual summarization of alerts, that has improved SOC analyst productivity. 
 However, these solutions are assistive tools and do not learn decision policies. 
 In contrast, the proposed framework offers a DRL–based policy layer that operates upstream of alert generation, learning cost-aware containment decisions over aggregated network traffic and identifying suspicious traffic window, and helps the SOC analyst in decision making to block or allow traffic in the network.
 As compared to monolithic copilot architectures, our framework uses a hierarchical agentic design in which DRL governs when and how LLM is invoked.
 Moreover, by invoking LLM only for high-priority events, our framework reduces computational overhead and analyst burden, positioning it as a complementary and orthogonal solution to existing copilot-based platforms.
\end{itemize}

Agentic AI presents both opportunities and challenges: On the one hand, it can automate threat identification and analysis, but their adoption also risks getting manipulated and requires careful human oversight to verify the information.
Previous works do not address the implementation agentic system without limiting auditability and analyst control.
Moreover, prior works lack a complete cycle from data ingestion into SIEM tool to anomalous flows findings, prioritization of flows, LLM based multi-agent alert triage, query generation to validation of IoCs on SIEM platform. 
Our work addresses this gap by integrating Agentic AI based analysis with SIEM platform such as Splunk~\cite{splunk} to provide automated, context-aware threat hunting, along with operational validation of threats on SIEM tool.  
Our multi-layered framework integrates the strengths of AI-driven analysis, LLM based contextual insights, robust SIEM capabilities, and providing a more effective and automated threat hunting framework.

\section{Key observations and design objectives}
\label{sec:key_observation}

We analyzed and compared the agentic AI and LLM based threat detection mechanisms in Section~\ref{sec:related_works}, and summarized in Table~\ref{tabl:key_observation}.
Our primary observations are highlighted below.
\begin{itemize}
    \item Most of the agentic AI based mechanisms in our survey are designed only for threat detection and do not integrate the SIEM platform into the framework.
    \item One of the desirable features is that Agentic AI based threat hunting framework should perform anomaly assessment and prioritize alerts and traffic flows before forwarding to LLM for contextual analysis to avoid processing overhead. 
    In~\cite{unifiedframeworkai}, the authors address both issues to some extent at the high-level.
    \item Initial triage is important for threat hunting before delving into more details.
    While some of the existing ones have this feature, but they provide this functionality at the level of LLM which causes processing overhead on the LLM due to the huge volume of information.
    \item Most of the mechanisms provide a decision support system for the SOC environment partially without any SOAR suggestion.
    \item While analyzing the traffic logs, it is also important to understand the attack technique used by the attackers.
    Therefore, the MITRE mapping is important to comprehend this.
    As we can see in Table~\ref{tabl:key_observation}, few methods address this issue. 
    \item As shown in Table~\ref{tabl:key_observation}, most of the mechanisms that we reviewed do not offer the feature to adapt according to the learned policy.
    Although few methods use autonomous adaptation to detect zero day attacks~\cite{autonoumousaiagent,agenticai-selfhealing}.
\end{itemize}

With these key observations in mind, we are motivated to design such an agentic AI based threat hunting framework that can overcome the identified limitations and achieve a number of desired properties, as follows:

\begin{itemize}
    \item Anomaly assessment and initial triage: Anomaly assessment, traffic prioritization, and initial triage are performed before forwarding the alert to LLM for analysis. 
    This closes the gap between detection and response with investigation support.
    \item SIEM Integration: The SIEM platform is integrated with the framework for further deep investigation of traffic flows.
    \item SOAR and SOC decision support: The framework provides SOAR suggestion and supports SOC analysts in decision making by assisting them in filtering traffic flows on the SIEM platform with queries. 
    \item MITRE Mapping: The mapping to MITRE ATT\&CK framework is done to understand the attack technique used. 
    \item Adaptive learned policy: The framework offers the learned policy layer to adapt according to the requirements of the SOC objectives.
    
\end{itemize}

Thanks to the aforementioned properties, the features ranging from anomaly assessment, initial triage, traffic prioritization, MITRE mapping and SIEM tool can be systematically integrated together, to assist the SOC analysts for informed decision making. 


\begin{table}[]
\centering
\caption{Comparison between existing LLM and Agentic AI based threat hunting techniques}
\label{tabl:key_observation}
\resizebox{\textwidth}{!}{%
\begin{tabular}{|l|l|l|l|l|l|l|l|l|l|}
\hline
\begin{tabular}[c]{@{}l@{}}Threat Hunting\\ Technique\end{tabular} & \begin{tabular}[c]{@{}l@{}}Anomaly \\ Assessment\end{tabular} & \begin{tabular}[c]{@{}l@{}}Traffic \\ Flows \\ Prioritization\end{tabular} & \begin{tabular}[c]{@{}l@{}}Initial \\ Triage\end{tabular} & \begin{tabular}[c]{@{}l@{}}Agentic AI/LLM \\ Integration\end{tabular} & \begin{tabular}[c]{@{}l@{}}MITRE\\  Mapping\end{tabular} & \begin{tabular}[c]{@{}l@{}}SIEM \\ Integration\\ and\\ Validation\end{tabular} & \begin{tabular}[c]{@{}l@{}}SOAR \\ Suggestion\end{tabular} & \begin{tabular}[c]{@{}l@{}}SOC\\ Decision\\ Support\end{tabular} & \begin{tabular}[c]{@{}l@{}}Learned Policy\\ adaptation\end{tabular} \\ \hline
\begin{tabular}[c]{@{}l@{}}Agentic AI for\\ Autonomous Threat\\ Hunting\\ ~\cite{autonoumousaiagent}\end{tabular} & Yes & No & No & Yes & No & No & No & Yes & Partial \\ \hline
\begin{tabular}[c]{@{}l@{}}Agentic AI for Adaptive\\ Threat Response\\ ~\cite{agenticai-selfhealing}\end{tabular} & Yes & No & No & Yes & No & No & No & \begin{tabular}[c]{@{}l@{}}Yes at\\ high-level\\ (Partial)\end{tabular} & Partial \\ \hline
\begin{tabular}[c]{@{}l@{}}Unified Framework for\\ Human-AI Collaboration\\ ~\cite{simbianwhitepaper,unifiedframeworkai}\end{tabular} & Conceptual & Partial & No & Yes & No & \begin{tabular}[c]{@{}l@{}}Yes at\\ high-level\end{tabular} & No & Yes & No \\ \hline
\begin{tabular}[c]{@{}l@{}}NetMon AI\\ Framework~\cite{netmonai}\end{tabular} & Yes & No & Yes & Yes & No & No & No & No & No \\ \hline
\begin{tabular}[c]{@{}l@{}}ARGOS Agentic\\ Detection~\cite{argos}\end{tabular} & Yes & No & No & Yes & No & No & No & No & No \\ \hline
\begin{tabular}[c]{@{}l@{}}LLM for Threat \\ Intelligence\\ and Automation\\ ~\cite{usingllmtoautomatectisoc}\end{tabular} & No & No & No & \begin{tabular}[c]{@{}l@{}}LLM \\ integration\end{tabular} & No & No & No & Partial & No \\ \hline
\begin{tabular}[c]{@{}l@{}}LLM for Non-Security\\  Experts\\ ~\cite{llmfornonsecurity}\end{tabular} & No & No & Yes & LLM Analysis & No & Yes & No & Partial & No \\ \hline
\begin{tabular}[c]{@{}l@{}}LLM Benchmarking\\ for Log Analysis\\ ~\cite{benchmarkingllm}\end{tabular} & No & No & Yes & \begin{tabular}[c]{@{}l@{}}LLM \\ Integration\end{tabular} & No & No & No & No & No \\ \hline
\begin{tabular}[c]{@{}l@{}}LLM for Security\\ alarm analysis\\ ~\cite{alarm_analysis_llm}\end{tabular} & No & No & Yes & \begin{tabular}[c]{@{}l@{}}LLM \\ Integration\end{tabular} & No & No & No & Partial & No \\ \hline
\begin{tabular}[c]{@{}l@{}}Microsoft Security\\ Copilot\end{tabular} & No & Partial & Yes & Yes & Partial & Yes & Yes & Yes & No \\ \hline
\begin{tabular}[c]{@{}l@{}}LLM and Splunk\\ for SOC~\cite{sahay_genai}\end{tabular} & No & No & Yes & LLM integration & Yes & Yes & Yes & Yes & No \\ \hline
\begin{tabular}[c]{@{}l@{}}Proposed \\ Framework\end{tabular} & Yes & Yes & Yes & Yes & Yes & Yes & Yes & Yes & Yes \\ \hline
\end{tabular}%
}
\end{table}

\begin{figure}[h]
    \includegraphics[width=\textwidth]{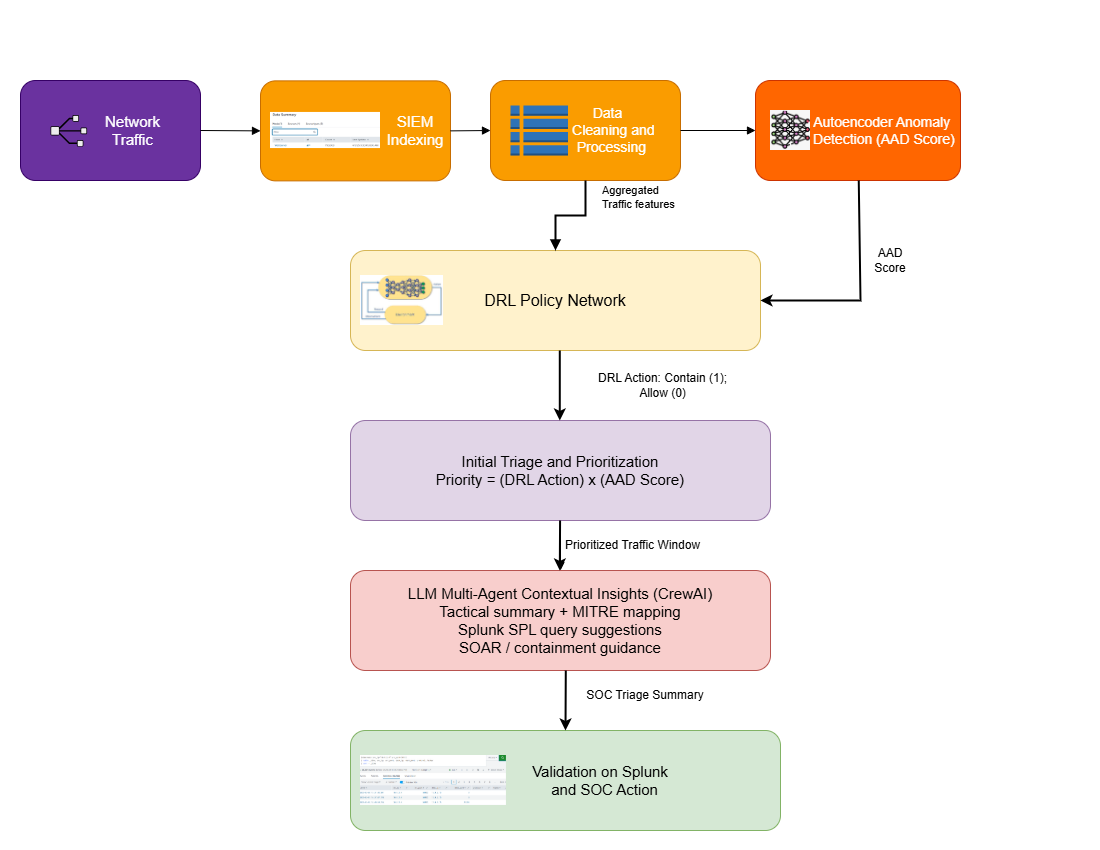}
    \centering
    \caption{Agentic AI-based Threat Hunting Framework}
    \label{doc:framework}
\end{figure}

The next Section~\ref{sec:core_components}, describes the major components of the proposed framework in detail.

\section{Agentic AI SOC Framework}
\label{sec:genai}

This section describes the major components of the proposed threat hunting framework.
As shown in Fig.~\ref{doc:framework}, the framework comprises of major components: SIEM Indexing, Data Cleaning \& Processing, Autoencoder Anomaly Detection, Deep Reinforcement Learning (DRL) Network, Prioritization, LLM Multi-agent triage, and Splunk validation.
The components are described in Section~\ref{sec:core_components}.

\subsection{Major Framework Components}
\label{sec:core_components}

\textbf{Data Collection and SIEM Indexing:} This component is responsible for indexing logs collected from different devices in the organization and store them in a central database for analysis. Collecting and indexing logs from different devices in the SIEM tool is an important step in threat hunting. 
The log collection agents are deployed on the devices that collect and forward the logs to the SIEM server.
It is also important to create a strategy to collect the volume of logs from different devices, as it can cause processing overhead on the SIEM tool as well as alert fatigue for the SOC analyst. 

\textbf{Data Cleaning and Processing:} This module is responsible for cleaning and  pre-processing the indexed data for further investigation.
All duplicate instances are removed because the data on the SIEM server often have many duplicates.
It extracts the features (such as IP addresses, ports, protocols, in bytes, out bytes, etc.) from the indexed data and exports them in csv format for processing. 

\textbf{Autoencoder-based Anomaly Detection:} In the framework, the reconstruction based deep autoencoder is trained on a part of legitimate traffic and learns the benign network pattern.
The autoencoder is a neural network that tries to reconstruct the normal network pattern and assigns the reconstruction error to data samples.
Data instances that are normal are reconstructed with minimal error.
However, anomalies are reconstructed with high error. 
In the framework, the reconstruction error is assigned to the data instances as the anomaly score (AAD). 
The AAD score along with the DRL decision is used to prioritize the malicious traffic window to be forwarded to LLM for contextual analysis.
The details are described in Section~\ref{sec:genai}.

\textbf{Deep Reinforcement Learning Network:} The DRL module in the framework comprises of three major components: a) DRL agent, b) simulation Environment, and c) Reward function.
The DRL agent interacts with the simulated environment that contains aggregated traffic features over a fixed time window, and takes action based on its current state.
The environment responds with a new state and a reward as feedback based on its action of the DRL agent.
The objective of the DRL agent is to take action in such a way as to maximize cumulative reward.
In the framework, action space also termed as the DRL decision comprises: a) Containment (1), b) Allow (0).

\textbf{Initial Triage and Prioritization:} Based on the DRL decision and the anomaly score (AAD Score) of the autoencoder, the \texttt{Initial Triage and Prioritization} module prioritizes malicious traffic windows for LLM analysis.
The DRL action (1 or 0) is multiplied by the AAD score to prioritize the flow.
This initial triage and prioritization of the malicious traffic window reduces the processing overhead on LLM.

\textbf{LLM Multi-Agent based Contextual Analysis:} This module analyzes the received malicious traffic window and provides the contextual insights.
It is based on a multi-agent system, where one LLM agent acts as an orchestrator and two other LLM agents perform the analysis.
In the framework, the agent titled \texttt{Senior SOC Triage Analyst} analyzes the traffic, provides its assessment, and generates SPL (Splunk Query Language) queries to filter the traffic on the Splunk dashboard. 
Another LLM agent called \texttt{Threat Intelligence Analyst} performs the mapping of traffic behavior to MITRE ATT\&CK framework. 
Finally, the orchestrator summarizes the contextual analysis in a human-readable format.
The listings~\ref{lst:triage_prompt} and~\ref{lst:threatintel_prompt} show the sample prompts for the SOC triage agent and the threat intelligence agent.
The implementation of this component is done using a CrewAI framework~\cite{crewai}.
Currently, in our proposed framework, public LLM' API such as ChatGPT is used for LLM triage, but local LLM is suggested for the analysis of sensitive logs.
Data anonymization can also be used to anonymize sensitive information before forwarding it to public LLM. 

\textbf{Validation in Splunk:} The insights provided by the LLM is further validated by the human SOC analyst on the Splunk dashboard by pulling the logs.
The SPL queries provided by the \texttt{Senior SOC Triage Analyst} is used by the human SOC analyst to validate the insights in Splunk.
Based on the findings in Splunk, the SOC analyst can decide to \texttt{Block} or \texttt{Allow} traffic.
For instance, if SOC analyst finds the presence of Indicators of Compromise (IoCs) such as a high number of packets in a very short time window in the logs, then they can block the traffic by configuring the detection rules.
It reduces the workload of the SOC analyst by directly identifying the relevant flows for investigation in Splunk and supplying the corresponding SPL queries.
It ensures that the LLM analysis is appropriately filtered.

\subsection{Design Rationale}

The key design principle in the proposed framework is strict functional separation to avoid feature leakage and model bias, as shown in Table~\ref{tabl:design_rationale}.
Specifically, the AAD score is explicitly excluded from DRL training to avoid leakage of features and inflated performance.
It is only included after DRL training for SOC triage and prioritization.

\begin{table}[h!]
\caption{Design rationale of the main components}
\centering
\label{tabl:design_rationale}
\begin{tabular}{|l|l|l|}
\hline
Component & Purpose                                                                                           & Input Feature                                                                                               \\ \hline
AAD       & \begin{tabular}[c]{@{}l@{}}Continuous anomaly \\ scoring\end{tabular}                             & Raw network features                                                                                        \\ \hline
DRL       & Decision, initial triage and prioritization                                                                       & \begin{tabular}[c]{@{}l@{}}Raw features except\\ AAD score\end{tabular}                                     \\ \hline
LLM Agent & \begin{tabular}[c]{@{}l@{}}Contextual insights and human\\ like triage and reasoning\end{tabular} & \begin{tabular}[c]{@{}l@{}}Prioritized traffic flow along\\ with DRL decision and AAD \\ score\end{tabular} \\ \hline
\end{tabular}
\end{table}

\newpage
\begin{lstlisting}[style=promptstyle,caption={A sample prompt for the SOC triage agent},label={lst:triage_prompt}]
Role: Senior SOC Triage Analyst
Goal: Assess the threat level associated with a prioritized network flow.
Backstory: Expert at distinguishing benign network noise from suspicious or malicious traffic.

Input:
- Flow ID: {flow_id}
- Source IP: {src_ip}
- Destination IP: {dest_ip}
- Destination Port: {dest_port}
- Priority Score: {priority}
- Anomaly Score: {aad_score}

Task:
Analyze the flow and determine whether the observed communication appears benign,
suspicious, or high risk. Briefly explain the reasoning using the provided network context.

Expected Output:
A concise SOC-style summary of the alert risk level.
\end{lstlisting}

\newpage
\begin{lstlisting}[style=promptstyle,caption={A sample prompt for the threat intelligence agent},label={lst:threatintel_prompt}]
Role: Threat Intelligence Analyst
Goal: Map suspicious activity to MITRE ATT&CK and provide remediation guidance.
Backstory: Expert in associating network behaviors with adversarial techniques and response actions

Input:
- Flow ID: {flow_id}
- Source IP: {src_ip}
- Destination IP: {dest_ip}
- Destination Port: {dest_port}
- Priority Score: {priority}
- Anomaly Score: {aad_score}

Task:
Identify the most relevant MITRE ATT&CK technique associated with the observed flow.
Provide the MITRE technique ID, technique name, and a brief remediation recommendation.

Expected Output:
MITRE ATT&CK technique ID, technique name, and remediation guidance.
\end{lstlisting}

\section{Methodology}
\label{sec:methodology}

In this section, we model the Security Operations Center (SOC) as a sequential decision-making problem, where a DRL agent is trained on aggregated network traffic features and decides whether to \texttt{allow} or \texttt{contain} the traffic. 
The system operates over a fixed-length, non-overlapping time windows and is formulated as a Markov Decision Process (MDP).
The DRL agent works exclusively on aggregated raw traffic features, while anomaly scores and LLM reasoning are introduced only after the containment decision is made.
The framework enforces strict separation of responsibilities to avoid feature leakage commonly seen in hybrid machine learning security systems.

\begin{algorithm}[h!]
\caption{Autoencoder-Based Anomaly Detection (AAD) with Benign-Only Standardization and Flow Mapping}
\label{alg:aad}
\begin{algorithmic}[1]

\Require
Flow dataset $\mathcal{D}$ with timestamps and identifiers (e.g., $\texttt{flow\_id}$), sorted by time;
feature subset $\mathcal{F}=\{\texttt{src\_port},\texttt{dest\_port},\texttt{bytes\_in},\texttt{bytes\_out}\}$;
benign label indicator $y \in \{0,1\}$ (used only for selecting benign training samples);
training fraction $\alpha$ (e.g., $0.25$);
autoencoder bottleneck dimension $d$ (e.g., $d=2$).

\Ensure
Anomaly score $\mathrm{AAD}(w_t)$ for each time window $w_t$ and a mapping $\mathcal{M}$ from each $w_t$ to its representative flow metadata.

\vspace{0.3em}
\State Sort $\mathcal{D}$ chronologically by timestamp.
\State Split $\mathcal{D}$ into an early training period $\mathcal{D}^{train}$ (first $\alpha$ fraction) and the full period $\mathcal{D}^{all}$.
\State Filter benign-only samples from the training period: $\mathcal{D}^{benign}=\{x \in \mathcal{D}^{train}\mid y(x)=0\}$.
\State Extract benign training matrix $X^{benign} \leftarrow \mathcal{D}^{benign}[\mathcal{F}]$ and full matrix $X^{all} \leftarrow \mathcal{D}^{all}[\mathcal{F}]$.

\vspace{0.3em}
\State \textbf{Benign-only standardization:}
\State Compute $\mu \leftarrow \mathrm{mean}(X^{benign})$ and $\sigma \leftarrow \mathrm{std}(X^{benign})$.
\State Standardize $X^{benign}_s \leftarrow (X^{benign}-\mu)/(\sigma+\epsilon)$ and $X^{all}_s \leftarrow (X^{all}-\mu)/(\sigma+\epsilon)$.

\vspace{0.3em}
\State Initialize reconstruction autoencoder $f_\phi$ with architecture $(|\mathcal{F}| \rightarrow 8 \rightarrow d \rightarrow 8 \rightarrow |\mathcal{F}|)$.
\State Train $f_\phi$ on benign data by minimizing reconstruction loss:
\[
\mathcal{L}(\phi)=\frac{1}{|X^{benign}_s|}\sum_{x \in X^{benign}_s} \|x - f_\phi(x)\|_2^2.
\]

\vspace{0.3em}
\State \textbf{Windowing and mapping:} Partition $\mathcal{D}^{all}$ into fixed windows $\{w_t\}$ (e.g., 5 minutes).
\For{each window $w_t$}
    \State Construct window feature vector $x_t \leftarrow \mathrm{Agg}(X^{all}_s \text{ within } w_t)$.
    \State Reconstruct $\hat{x}_t \leftarrow f_\phi(x_t)$.
    \State Compute anomaly score:
    \[
    \mathrm{AAD}(w_t)=\frac{1}{|\mathcal{F}|}\sum_{i \in \mathcal{F}}(x_{t,i}-\hat{x}_{t,i})^2.
    \]
    \State Store mapping $\mathcal{M}(w_t)$ to contextual metadata (e.g., $\texttt{src\_ip}$, $\texttt{dest\_ip}$, $\texttt{dest\_port}$, $\texttt{flow\_id}$) using representative values such as \texttt{first} or \texttt{mode} within $w_t$.
\EndFor

\State \Return $\{\mathrm{AAD}(w_t)\}$ and mapping $\mathcal{M}$.

\end{algorithmic}
\end{algorithm}

\subsection{Autoencoder-Based Anomaly Detection (AAD)}








In our framework, the anomaly score (AAD score) is computed  using a fully connected reconstruction based autoencoder anomaly detection (AAD) with a low dimensional bottleneck architecture (8-2-8), which is trained on early benign traffic flows to model normal network behavior~\cite{deeplearningaad}.
The Algorithm~\ref{alg:aad} implements an unsupervised Autoencoder trained on legitimate traffic observed during an initial period (we use the first fraction of the timeline, e.g., 25\%).
Feature normalization is performed using statistics derived only from early benign traffic flows to simulate a clean baseline of normal network behavior to avoid mixing the anomaly model with attack information. 
Given a standardized feature vector $x_t \in \mathbb{R}^{|\mathcal{F}|}$, the autoencoder reconstructs $\hat{x}_t$, and computes the anomaly score as the mean squared reconstruction error shown in Equation~\ref{eq:aad_score}.
The anomaly score is computed per flow and then aggregated.

\begin{equation}
\label{eq:aad_score}
\mathrm{AAD}(t)=\frac{1}{|\mathcal{F}|}\sum_i (x_{t,i}-\hat{x}_{t,i})^2.    
\end{equation}

As shown in Algorithm~\ref{alg:aad}, the model first learns a compressed latent representation of normal network features and assigns a higher reconstruction error to malicious flows.
In our framework, the AAD Score is not used as part of the DRL state representation during policy learning; instead, it is utilized after DRL decisions to prioritize DRL flagged traffic flows for LLM triage.
Although the autoencoder is trained only on numeric features (ports and byte counts), the anomaly score is not detached from flow context.
Each score is computed per fixed time window $w_t$ and is stored together with a metadata mapping $\mathcal{M}(w_t)$ containing representative identifiers (e.g., \texttt{flow\_id}, \texttt{src\_ip}, \texttt{dest\_ip}, and service port).
In practice, the autoencoder produces $\mathrm{AAD}(w_t)$ from $\mathcal{F}$, while attribution fields are preserved outside the model and re-attached to the scored window for downstream DRL prioritization and LLM triage.
This design prevents leakage from high-cardinality categorical fields while maintaining interpretability for SOC analysts.

\begin{table}[h!]
\caption{Reconstruction Autoencoder vs Classification Model}
\label{tabl:autoencoder_classification}
\centering
\begin{tabular}{|l|l|l|}
\hline
Aspect          & State Reconstruction Autoencoder & General Classification Model \\ \hline
Labels required & No                               & Yes                          \\ \hline
Learns          & Normal behavior                  & Decision boundary            \\ \hline
Output          & Anomaly score                    & Discrete class               \\ \hline
Used for        & Prioritization, Triage           & Detection                    \\ \hline
\end{tabular}
\end{table}

In the framework, the reconstruction based autoencoder is different from general classification models.
The reconstruction based autoencoder used in the framework learns normal benign traffic feature and provides anomaly score (AAD).
To learn the traffic behavior it does not need any labels.
This anomaly score is used for traffic prioritization for triage.
However, general classification models, as shown in Table~\ref{tabl:autoencoder_classification} require labels and provide a discrete class as a decision.
It is generally used for threat detection.

\subsection{Time-Window State Construction}

Raw network traffic is represented as a sequence of flow records, each comprising of source \& destination addresses, ports, protocol, in bytes and out bytes.
Traffic is aggregated into fixed-length windows of duration $\Delta t = 5$ minutes.
For each time window $t$, aggregated state vector is represented by $s_t$ by merging numerical statistics (such as mean and maximum port numbers and byte counts) along with encoded categorical attributes such as protocol identifiers and low-cardinality representations of IP addresses.
The use of fixed temporal window is widely considered best practices in modern Security Operations Centers~\cite{splunk_correlation_searches}.

For example, let $F_t = \{f_1, f_2, \dots, f_{N_t}\}$ denote the set of raw network events observed during time window $t$, where $N_t$ may be on the order of millions in high-throughput SOC environments.
Modeling the state of the system as $s_t = F_t$ is computationally infeasible.
Instead, we build a state representation using an aggregation function represented by $\phi(\cdot)$; where $\phi(\cdot)$ represents a large set of raw traffic in a fixed-dimensional feature vector that captures traffic statistics, protocol behavior, and temporal characteristics.
\begin{equation}
s_t = \phi(F_t)    
\label{eq:aggregation}
\end{equation}

This aggregation represents both regular and burst-oriented patterns while simultaneously reducing noise and dimensionality.
Detailed feature engineering, aggregation equations and examples are provided in Appendix~A. 

\subsection{Deep Reinforcement Learning for SOC}
\label{sec:drl_soc}

We employ Deep Reinforcement Learning (DRL) to build our containment agent, often referred to as a DRL agent in the paper, which directly learns from traffic features to identify anomalies.
Specifically, we model the containment decision problem as a sequential decision-making process using deep reinforcement learning (DRL) by defining the state and action spaces for the agent.
Formally, the containment decision is modeled as a Markov Decision Process (MDP)
$\mathcal{M} = (\mathcal{S}, \mathcal{A}, \mathcal{P}, \mathcal{R}, \gamma)$,
where $\mathcal{S}$ denotes the state, $\mathcal{A}$ represents the action space, $\mathcal{P}$ represents state transitions across time windows, $\mathcal{R}$ is a reward function reflecting SOC priorities, and $\gamma \in (0,1]$ is the discount factor.

The state of the DRL agent at the time step $t$, is represented by $s_t$, that signifies the aggregated network traffic over a fixed time window. 
Equation~\ref{eq:aggregation} represents the state of the system over aggregated traffic features.
The action space of the DRL agent is defined as $\mathcal{A} = \{0,1\}$.
At each time step $t$, given the state $s_t$, the DRL agent selects an action $a_t = 1$, if traffic $F_t$ is considered malicious and $a_t = 0$ if $F_t$ is assessed as legitimate. 
The aim of the agent is to learn an optimal containment policy that maximizes long-term operational utility while balancing detection accuracy and alert fatigue.

The DRL policy network represented as $\pi_\theta(a_t \mid s_t)$, is implemented using Multilayer Perceptron (MLP) i.e., a feedforward neural network with two hidden layers of 64 neurons each, as mentioned in Algorithm~\ref{alg:drl_soc}.
The network takes the aggregated traffic state $s_t \in \mathbb{R}^d$ as input and outputs a probability distribution over two actions: \emph{Contain} and \emph{Allow}.
The $s_t \in \mathbb{R}^d$ means that the state at time $t$ is a real-valued vector of dimension $d$.
In our framework, we use Rectified Linear Unit (ReLU) activations in hidden layers, and a softmax function is applied at the output layer to produce action probabilities~\cite{relu}.


Given the state $s_t$, the forward pass of the policy network is defined as follows.

\paragraph{Hidden Layer 1}
\begin{equation}
h_1 = \mathrm{ReLU}(W_1 s_t + b_1),
\end{equation}

\paragraph{Hidden Layer 2}
\begin{equation}
h_2 = \mathrm{ReLU}(W_2 h_1 + b_2),
\end{equation}

\paragraph{Output Layer (Logits)}
\begin{equation}
z = W_3 h_2 + b_3,
\end{equation}
where $z \in \mathbb{R}^2$ contains unnormalized action scores (logits) for \emph{allow} and \emph{containment}.
The unnormalized action scores are raw outputs of the policy neural network that signifies how the agent favors each action before softmax normalization.

The action probabilities are calculated using softmax and the confidence of the action \texttt{contain} is given by $\max_a \pi(a \mid s_t)$.
\begin{equation}
\pi(a \mid s_t) = \frac{\exp(z_a)}{\sum_{a' \in \{0,1\}} \exp(z_{a'})}.
\end{equation}


The anomaly score provided by the Autoencoder-based Anomaly Detection (AAD) to the traffic flow is not included in the DRL state representation.
It prevents feature leakage and make sure that the DRL agent learns containment behavior exclusively from raw and aggregated network traffic, rather than precomputed anomaly scores.
The anomaly score is used after DRL decision while prioritizing the traffic flow for LLM triage. 

For each time window, a triage priority score is computed as shown in Equation~\ref{eq:triage}.
\begin{equation}
\label{eq:triage}
\text{Triage Priority} = \text{DRL\_Action} \times \text{AAD\_Score}
\end{equation}

This formulation ensures that only windows flagged by the DRL agent are prioritized, while the anomaly score validates their urgency.
Highly anomalous flows receive higher priority,
while low-risk anomalies are de-prioritized even if flagged for containment.
This mirrors real SOC analyst workflows, where decisions are discrete but prioritization is continuous.

The policy optimization is performed using Proximal Policy Optimization (PPO), which stabilizes learning by constraining policy updates within a clipped trust region~\citep{ppo}.
Temporal consistency is preserved through time-series cross-validation, ensuring that training data strictly precede test data in the timeline.
The agent iteratively observes the current network state, selects an action according to its policy, and receives a scalar reward based on the correctness of the decision relative to ground truth labels available during training.
The multiple reward policies are described in Section~\ref{sec:reward_shaping}.

In summary, PPO decides how the policy is updated in the framework.
The MLP based on $2 \times 64$ i.e. 2 layer and 64 neurons shows what the policy looks like.
Together, they build DRL agent that:
\begin{itemize}
    \item Observes aggregated network traffic.
    \item Predicts an action (contain or allow).
    \item Receives a reward based on correctness and cost.
    \item Updates the policy in a stable and constrained manner.
\end{itemize}

\begin{algorithm}[t]
\caption{Deep Reinforcement Learning for SOC Containment}
\label{alg:drl_soc}
\begin{algorithmic}[1]

\Require
Aggregated network windows $\mathcal{W} = \{w_1, w_2, \dots, w_T\}$ with feature vectors $s_t$  
Ground-truth labels $y_t$ (used only for reward computation during training)

\Ensure
Containment policy $\pi_\theta(a \mid s)$

\vspace{0.3em}
\State Initialize PPO policy network $\pi_\theta$ with 2 hidden layers (64 neurons each)
\State Define action space $\mathcal{A} = \{0:\text{No Action}, 1:\text{Containment}\}$
\State Define reward function $\mathcal{R}(a_t, y_t)$ emphasizing low false positives

\vspace{0.3em}
\For{each training episode}
    \For{each time step $t$}
        \State Observe environment state $s_t$
        \State Sample action $a_t \sim \pi_\theta(a_t \mid s_t)$
        \State Execute action $a_t$
        \State Receive reward:
            $r_t = \mathcal{R}(a_t, y_t)$
        \State Store $(s_t, a_t, r_t)$
    \EndFor
    \State Update policy parameters $\theta$ using PPO objective
\EndFor

\vspace{0.3em}
\State \Return trained containment policy $\pi_\theta$

\end{algorithmic}
\end{algorithm}


\subsubsection{Reward Shaping}
\label{sec:reward_shaping}

The reward function is designed to reflect SOC operational priorities, with a particular emphasis on minimizing false positives, which are costly and disruptive in real-world environments.
Correct containment of malicious activity (true positives) is positively rewarded, while false positives incur significant penalties.
True negatives are rewarded to reinforce restraint (no containment) when traffic is benign, while false negatives are penalized to discourage missed detections.
Multiple reward profiles are evaluated to study the trade-off between detection sensitivity and false-positive reduction.

\begin{equation}
r_t = \mathcal{R}(a_t, y_t)
\label{eq:reward_general}
\end{equation}

In the proposed DRL formulation, the agent operates in a binary decision space at each time step \(t\).
Let \(a_t \in \{0,1\}\) represent the action selected by the agent, where ($a_t = 1$) corresponds to containment decision (e.g., blocking or isolating a network flow) and \(a_t = 0\) represents \texttt{allow} action i.e., traffic is benign.
The ground-truth label at time step \(t\) is shown by \(y_t \in \{0,1\}\), where \(y_t = 1\) denotes that the observed network flow is malicious, and ($y_t = 0$) shows benign traffic.

The agent’s action \(a_t\) and the true label \(y_t\) result in one of four security outcomes: true positive (TP), false positive (FP), false negative (FN) and true negative (TN).
These outcomes are the basis for the reward shaping strategy used to train the deep reinforcement learning agent.
By modeling this interaction, the agent learns policies that balance detection effectiveness with operational costs such as alert fatigue and unnecessary containment actions.

To study the impact of reward shaping on decision-making behavior, we define four distinct reward profiles (Modes A--D).
Each profile represents a different operational focus that is typically found in Security Operations Center (SOC) environments.

\textbf{Mode A} is a recall-oriented detection strategy.
In this mode, both true positives and true negatives are rewarded equally, while false negatives are penalized more than false positives.
It motivates the agent to prioritize detection coverage and early identification of malicious activity, making Mode A appropriate for initial threat discovery and baseline sensitivity analysis rather than strict false-positive control.

\textbf{Mode B} focuses on false positive reduction.
In this mode, false positives are penalized more heavily than false negatives, while true negatives are assigned a significant positive reward.
This reflects a SOC environment in which excessive containment measures and alert fatigue can incur significant costs.
Specifically, it is effective in prioritizing high-confidence alerts and reducing workload on SOC analysts.

\textbf{Mode C} provides a moderate trade-off between detection and operational cost.
False positives and false negatives are penalized at intermediate levels, while correct decisions receive moderate rewards.
This policy indicates SOC environments where both false positives and false negatives are considered bad, but neither dominates the operational policy.

\textbf{Mode D} incorporates controlled randomness into the reward function.
It preserves a balanced reward scheme related to Mode C, but adds Gaussian noise onto the reward signal.
This form of stochastic regularization improves policy robustness by reducing overfitting to fixed reward structures and by mimicking the uncertainty found in the SOC environment and analyst responses. 
It is represented by Equation~\ref{eq:reward_modeD_noise}.

\begin{equation}
r_t^{(D)} = \mathcal{R}(a_t, y_t) + \epsilon_t,
\qquad
\epsilon_t \sim \mathcal{N}(0,\sigma^2).
\label{eq:reward_modeD_noise}
\end{equation}

\subsubsection{DRL Objective}
\label{subsec:rl_objective}

The agent learns a containment policy $\pi_\theta(a_t\mid s_t)$ that maps the aggregated flow state $s_t$ to an action space over $(0,1)$.
The learning objective is to maximize the expected discounted return shown in Equation~\ref{eq:rl_objective}.

\begin{equation}
J(\theta)=\mathbb{E}_{\pi_\theta}\left[\sum_{t=0}^{T}\gamma^t r_t\right],
\label{eq:rl_objective}
\end{equation}

Where $\gamma \in (0,1]$ is the discount factor and $T$ is the episode horizon. 
Since the reward $\mathcal{R}(a_t,y_t)$ is designed to reflect the operational priorities of SOC, maximizing $J(\theta)$ is
equivalent to minimizing the long-term operational burden under the chosen reward profile.

\subsubsection{Decision Cost and Regret Analysis}
\label{subsec:cost_regret_drl}


To complement the standard detection metrics, we evaluate the learned containment policy using decision cost and regret, that provide insight into the operational quality of the agent’s actions under SOC constraints.

The decision cost at time step $t$ is defined as the negative of the obtained reward shown in Equation~\ref{eq:cost}:
\begin{equation}
\label{eq:cost}
\mathrm{cost}_t = -r_t.
\end{equation}

This formulation reflects the operational burden associated with an action.
High costs relates to undesirable outcomes, such as unnecessary containment of benign traffic or missed detection of malicious activity, while low or negative costs show decisions aligned with SOC priorities.

While decision cost evaluates the absolute penalty of an action, regret measures how far the agent’s decision deviates from the best possible decision at that time step.
Equation~\ref{eq:regret} defines the regret
.
\begin{equation}
\label{eq:regret}
\mathrm{regret}_t = r_t^\star - r_t,
\end{equation}
where
\begin{equation}
r_t^\star = \max_{a \in \{0,1\}} \mathcal{R}(a, y_t)
\end{equation}
shows the maximum achievable reward assuming perfect knowledge of the ground-truth label $y_t$.

By reporting both decision cost and regret, we measure not only whether the agent makes correct decisions, but also how costly its mistakes are in practice.
This dual perspective is important in SOC environments, where different
errors have different operational consequences.

\subsubsection{Putting it Together}

Unlike traditional supervised intrusion detection systems, the DRL agent in the framework does not aim to maximize classification alone.
Instead, it learns a policy optimized for long-term operational efficiency under SOC constraints.
The DRL agent acts as a decision making filter that determines whether an alert needs escalation reducing the analyst workload.

Specifically, the DRL agent is trained without using anomaly detection scores (e.g., AAD Score) as input features to avoid feature leakage; the anomaly scores are computed separately and only applied after DRL to prioritize traffic flows for LLM-based triage.
This separation ensures that the DRL policy generalizes beyond specific anomaly models and remains robust to changes in downstream scoring mechanisms.

\subsection{LLM-Based Multi-Agent SOC Triage}
\label{sec:llm_triage}

\begin{algorithm}[t]
\caption{LLM-Based Multi-Agent SOC Triage}
\label{alg:llm_soc}
\begin{algorithmic}[1]

\Require
Flagged traffic flows $\mathcal{F}$ from DRL agent,  
Anomaly scores $\text{AAD}(w_t)$, (Traffic flows with triage score > 5)  
Contextual metadata (IPs, ports, timestamps)

\Ensure
SOC triage reports with MITRE mapping and remediation guidance

\vspace{0.3em}
\For{each traffic flow $f_t \in \mathcal{H}$}
    \State Construct contextual prompt:
    \[
        \langle s_t, a_t, \text{AAD}(w_t), \text{network metadata} \rangle
    \]
    \State Invoke \textbf{SOC Analyst Agent}:
    \State \quad Generate tactical summary and risk assessment
    \State \quad Generate SPL query to investigate in Splunk 

    \State Invoke \textbf{Threat Intelligence Agent}:
    \State \quad Map activity to MITRE ATT\&CK techniques
    \State \quad Suggest SOAR or mitigation actions

    \State Store structured triage report
\EndFor
\vspace{0.3em}
\State \Return SOC investigation reports and master triage table

\end{algorithmic}
\end{algorithm}

The Algorithm~\ref{alg:llm_soc} highlights triage using Large Language Models.
Traffic flows with a priority score greater than 5 assigned after DRL decisions are forwarded to LLM for analysis.
Once a traffic window is prioritized by the DRL agent, individual flows within that window are extracted and forwarded for the LLM-based analysis.
The LLM operates at flow-level granularity, generating per-flow contextual insights, MITRE ATT\&CK mappings, and recommendations for the SOC analysts.
This design enables scalable decision-making by first filtering high-risk windows and performing detailed analysis only on selected flows.

Traffic flows forwarded to the LLM with contextual metadata, including source and destination addresses, ports, time windows, and the corresponding AAD scores.
The source and destination IP addresses are anonymized before being forwarded to the LLM for analysis~\cite{aiaugmentedsoc}.

Large Language Model based triage is implemented using a CrewAI multi-agent framework~\cite{crewai}. 
During the analysis process, two specialized agents are instantiated:
\begin{itemize}
    \item a Senior SOC Triage Analyst in the framework validates containment actions, assesses immediate risk, provides SPL (Splunk Processing Language) queries to pull the traffic flows on Splunk
    \item a Threat Intelligence Analyst maps the observed behavior to MITRE ATT\&CK techniques and recommends mitigation strategies. 
\end{itemize}

LLM agents with specialized SOC analyst roles provide human-readable summaries with contextual insights. 
This stage transforms low-level traffic flow information into actionable intelligence, reducing the workload of the SOC analyst, and accelerating decision making.
Moreover, SPL (Splunk Processing Language) queries generated by the Senior SOC Triage Analysts agent can be used by analysts to filter the logs on the Splunk dashboard, helping SOC analysts to filter the anomalous flows from the huge volume of logs. 
In the framework, the DRL agent operates at the traffic window level to identify high-risk traffic windows, therefore acting as a policy-level filter that reduces the volume of data processed by the DRL agent for investigation.
Once traffic windows are prioritized, individual flows within that window are extracted and analyzed by the LLM agents at the per-flow granularity.
Finally, the outputs are automatically compiled into per-flow investigation reports and a master SOC triage summary, enabling direct integration with Security Orchestration, Automation, and Response (SOAR) platforms.

In the framework, public LLM such as ChatGPT is used for the initial prototype.
However, local large language models can also be used.
So, to preserve data confidentiality, sensitive information such as IP addresses is anonymized prior to LLM-based analysis using deterministic pseudonymization~\cite{pseudonymization}.
It replaces real identifiers with stable tokens, ensuring consistency across time windows while preventing the exposure of sensitive information to public LLM.
The anonymization is non-destructive and reversible only within the SOC-controlled environment.

LLM agents generate SPL queries using anonymized identifiers which function as placeholders rather than executable values.
Prior to execution on Splunk, these placeholders are resolved through mapping the anonymized tokens back to their original identifiers that are maintained in a table.
As a result, the generated SPL queries remain applicable to the SIEM platform while ensuring that original information are never exposed to the LLM.





\section{Use Case}
\label{sec:use_case}

\begin{figure}[t!]
    \includegraphics[width=0.7\textwidth]{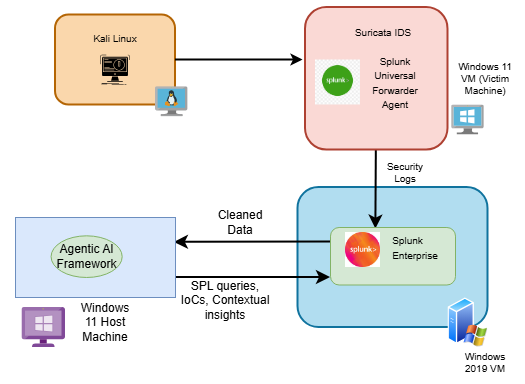}
    \centering
    \caption{Use case illustrating the application of framework}
    \label{doc:topology}
\end{figure}

For better understanding of the workflow, we provide a use case, as shown in Fig.~\ref{doc:topology}, in which the log collection agent is deployed on \texttt{Windows 11} running inside the virtual machine.
As shown in Fig.~\ref{doc:topology}, Kali Linux is used as an attacker machine to launch the attack on \texttt{Windows 11} that is the target.
The \texttt{Splunk Universal Forwarder} deployed on \texttt{Windows 11} collects and forwards logs to the \texttt{Splunk Enterprise} running on the \texttt{Windows 2019} server.
For the experimental evaluation, logs of Suricata IDS is forwarded by the \texttt{Splunk Universal Forwarder} to \texttt{Splunk Enterprise}.
The received logs are indexed on the \texttt{Splunk Enterprise} for investigation.
However, we plan to extend the use case with multiple log sources.
As shown in Fig.~\ref{doc:topology}, the Suricata IDS is deployed on \texttt{Windows 11}.

Once the logs are indexed, data cleaning and aggregation is applied 
before further processing using autoencoder and DRL.
For the autoencoder, a benign sample is filtered from an early temporal window, ensuring that the learned representation shows normal network behavior without mixing with attack traffic.
Moreover, alert labels and non behavioral features such as hashes are removed.
Furthermore, for the DRL training, a window based aggregation and cardinality reduction is performed to make sure that the DRL agent operates on the non leaking raw events. 
After DRL training, the traffic window is prioritized for LLM triage, using the DRL decision and anomaly score (AAD score).
Then, CrewAI based LLM agents analyze the received traffic window and provide contextual insights along with the SPL queries for further validation on Splunk for decision making, as described in Section~\ref{sec:genai}.
Based on the insight provided by the Agentic AI framework, entities such as IP addresses and ports are extracted and analyzed in the \texttt{Splunk Enterprise} by the SOC analyst for effective decision making.
The analysis performed using Splunk is shown in Section~\ref{sec:experimentation_results}.
During the evaluation with the public dataset (Boss of the SOC), we directly ingested the dataset into Splunk Enterprise and then exported it into CSV format for evaluation using autoencoder and the DRL.
The detailed results are discussed in Section~\ref{sec:experimentation_results}.

\subsection{Threat Model for Simulation}
\label{sec:threat_model}

\begin{itemize}
    \item \textbf{Network Scanning Attack:}It aims to identify active hosts on a network along with the ports and services running on the hosts\citep{networkscan}.
    It helps attackers assess the weaknesses of assets and plan the attack to exploit those vulnerabilities~\cite{sahay_genai}.
    \item \textbf{Volumetric Denial of Service Attack (DoS):} A volumetric DoS attack such as a UDP flood targets the victim system and the network to deplete resources such as network bandwidth, CPU, and memory by sending a large number of bogus packets\citep{Sahay2015TowardsAD}.
\end{itemize}

\section{Experimentation Results}
\label{sec:experimentation_results}

This section analyzes the experimentation results on the Boss of the SOC~\cite{bossofthesoc} and the simulated dataset.
We only used a part of the BOSS of the SOC dataset for this evaluation.
After cleaning and removing duplicates we used around 12000 instances.
The dataset contains source IP, destination IP, source port, destination port, in bytes, out bytes, protocols, and time.
We also simulated the use case described in Section~\ref{sec:use_case} and collected the dataset for the evaluation.
The simulated dataset also contains the same features as in the public dataset for evaluation and a total of 300000 instances.

\subsection{Evaluation of Boss of the SOC dataset}
\label{sec:boss_of_soc_evaluation}

Table~\ref{tabl:llm_triage_drl} shows the summary of the prioritized traffic flows forwarded to LLM for analysis.
Each traffic flow is characterized by its flow identifier, source and destination IP addresses, assigned priority score by the DRL agent after training.
The corresponding MITRE ATT\&CK mapping and the final conclusion is provided by the \texttt{Threat Intelligence Analyst} and the \texttt{Senior SOC Triage} agents. 
As we can see in Table~\ref{tabl:llm_triage_drl}, the flow (Flow ID:26) with the priority score of ($7.12 \times 10^{12}$) is mapped to technique T1071 (Application Layer Protocol), which is commonly associated with command-and-control (C2) communication channels.
The LLM agent classified this flow as \emph{critical}, highlighting a high-confidence detection of malicious command and control activity that requires immediate containment. 
This indicates that our framework segregates high-impact threats from background traffic to facilitate quick response from the SOC analyst.

\begin{table}[]
\caption{Automated LLM Triage based on Adaptive Scoring and Reinforcement Learning Agent}
\label{tabl:llm_triage_drl}
\begin{tabular}{|l|l|l|l|l|l|}
\hline
Flow ID & Source IP     & Destination IP  & Priority Score & MITRE ID  & Agent Answer                                                                                          \\ \hline
26      & 172.31.38.181 & 172.31.0.2      & 7.12e+12       & T1071     & \begin{tabular}[c]{@{}l@{}}Critical: C2 Communication\\  identified\end{tabular}                             \\ \hline
35      & 172.16.0.178  & 172.16.3.197    & 9.49e-01       & T1071     & \begin{tabular}[c]{@{}l@{}}Malicious behavior: Suspected\\  exfiltration via standard protocols\end{tabular} \\ \hline
34      & 192.168.8.103 & 192.168.9.30    & 7.33e-01       & T1071     & \begin{tabular}[c]{@{}l@{}}Risk level medium. Continuous\\ monitoring is advised.\end{tabular}               \\ \hline
30      & 172.16.0.178  & 169.254.169.254 & 5.22e-01       & T1552.005 & \begin{tabular}[c]{@{}l@{}}Possible attempt to exploit\\ cloud infrastructure\end{tabular}                   \\ \hline
23      & 192.168.8.112 & 192.168.9.30    & 5.06e-01       & T1071     & \begin{tabular}[c]{@{}l@{}}A moderate level of\\ concern. Continuous monitoring\\ advised\end{tabular}       \\ \hline
\end{tabular}
\end{table}

\subsection{SOC Analyst Validation using Splunk Analysis on Public Dataset}
\label{sec:splunk_analysis}

\begin{figure}[h]
    \includegraphics[width=\textwidth]{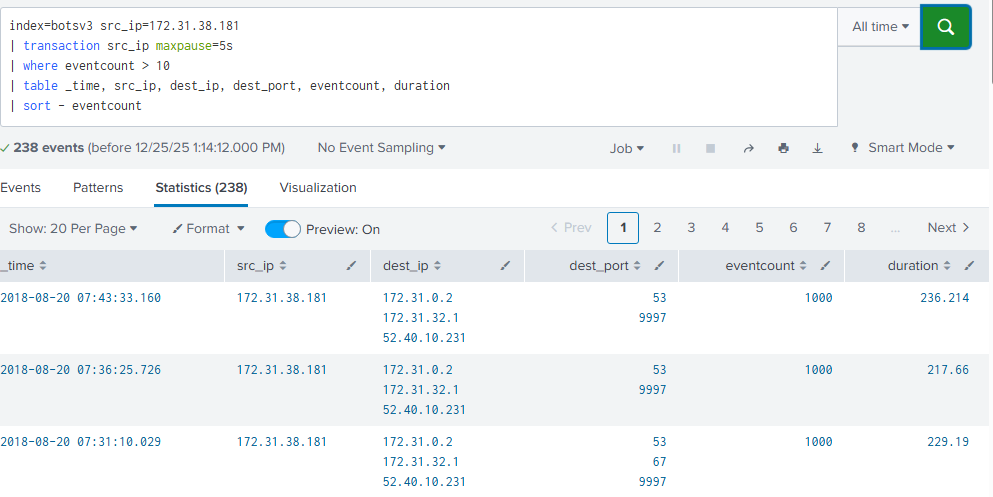}
    \centering
    \caption{Analysis of DNS traffic identified by the proposed RL and AAD triage mechanism}
    \label{doc:dns_burst}
\end{figure}


After prioritizing suspicious traffic using the proposed DRL and AAD framework, an LLM-based triage module performed contextual analysis and generated Splunk queries for the SOC analyst to validate the alert context within a real SOC environment. 
Fig.~\ref{doc:dns_burst} shows the SPL query applied to filter traffic originating from the specific host (src\_ip = 172.31.38.181) for the analysis.
Specifically this SPL query is very useful in network security analysis.
As shown in Fig.~\ref{doc:dns_burst}, the \texttt{transaction} operator groups individual network events into logical communication sessions based on a temporal threshold (maxpause = 5 seconds), reconstructing burst-oriented pattern.
The analysis suppresses benign, sporadic activity and isolates automated communication patterns by filtering events with \texttt{event count> 10}.

As we can see  in Fig.~\ref{doc:dns_burst}, a continuous high-frequency DNS communication (dest\_port = 53) originating from the IP address 172.31.38.181, with 1000 events occurring within a single transaction window lasting between 217 and 236 seconds. 
This continuous DNS activity is inconsistent with normal user behavior and is commonly found with command-and-control beaconing or DNS-based tunneling techniques.
Furthermore, the presence of multiple destination IP addresses within the same window highlights repeated attempts rather than a legitimate service communication.
The destination port 9997 shown in Fig.~\ref{doc:dns_burst} is used by the Splunk universal forwarder to send data to Splunk Enterprise.

Additionally, it is found that all DNS responses are syntactically valid (e.g., \texttt{reply\_code = NoError}), highlighting that detection is driven by behavioral aggregation rather than protocol violations or signature matching.
The Splunk analysis confirms the correctness of the DRL and AAD framework and demonstrates that the LLM-assisted triage provides quick validation of anomalous network pattern and helps the SOC analyst in informed decision making.


\subsection{Policy Adaptation Across Reward Modes with Boss of the SOC Dataset}


\begin{figure}[h]
    \includegraphics[width=\textwidth]{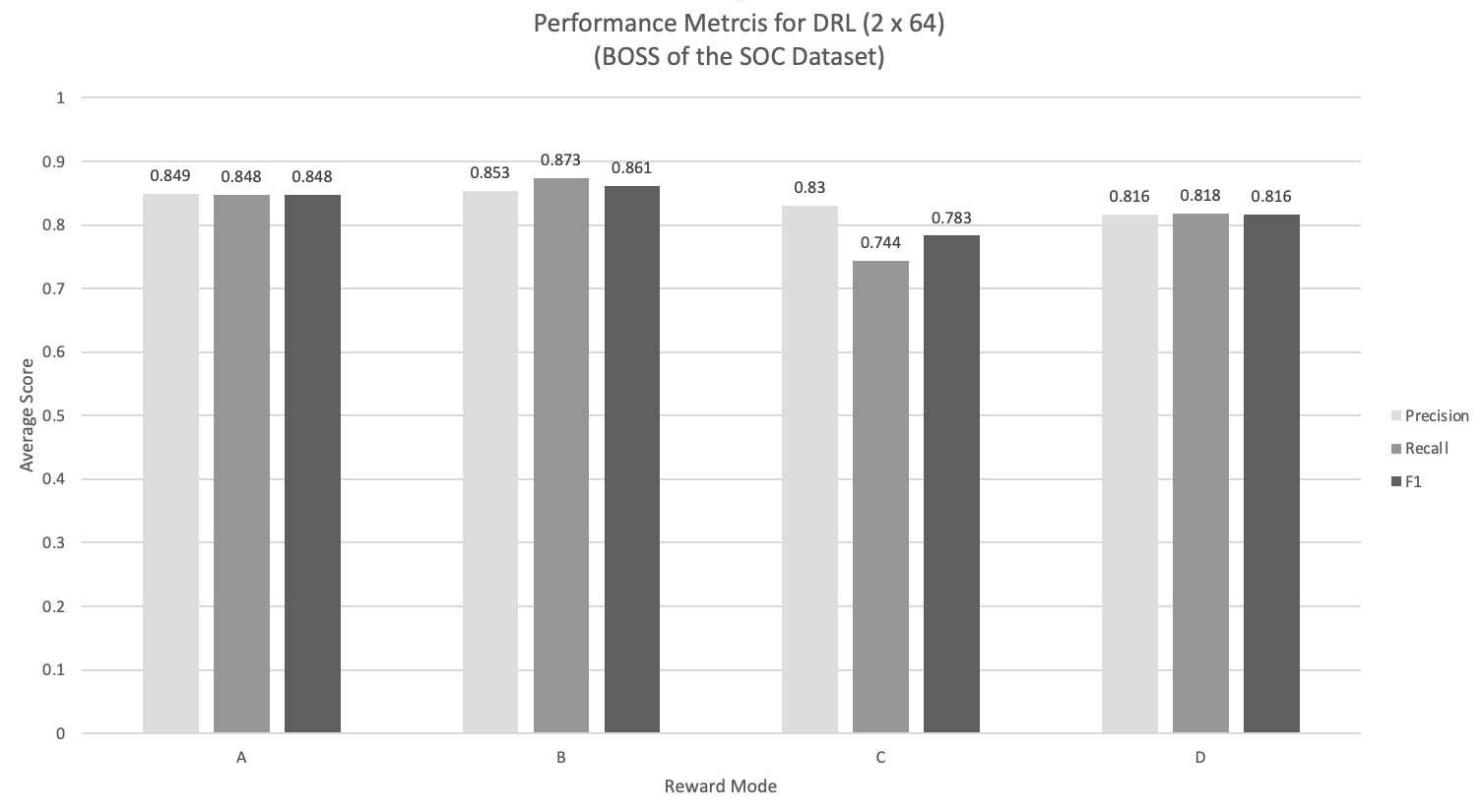}
    \centering
    \caption{Performance Evaluation of Across Modes (A-D)}
    \label{doc:metrics_bots}
\end{figure}

Fig.~\ref{doc:metrics_bots} presents the performance of the proposed two layer DRL-based agent across the four reward modes on the BOSS of the SOC dataset~\cite{bossofthesoc}.
As we can see in Fig.~\ref{doc:metrics_bots}, reward shaping directly impacts the trade-off between precision, recall, and overall detection effectiveness.

Mode A, which focuses on recall by penalizing false negatives more than false positives, achieves balanced performance with precision, recall, and F1-scores close to 0.85.
This highlights that the agent maintains a stable detection capability while allowing a limited number of false positives, making it suitable for early-stage threat detection. 

As shown in Fig.~\ref{doc:metrics_bots}, Mode B provides the best overall results, achieving the highest recall (0.873) and F1-score (0.861).
It is because of the reward function that rewards true negatives and penalizes false positives severely, enabling the agent to identify malicious activity with high confidence while limiting containment actions that are not required.

As we can see in Fig.~\ref{doc:metrics_bots}, Mode C shows a notable drop in recall (0.744) and the F1-score (0.783), highlighting the impact of balanced but strict penalties on both false positives and false negatives. 
Although precision remains comparatively high (0.830), the reduced recall indicates a more conservative policy that reduces alerts at the cost of missing some malicious events.

Finally, Mode D shows stable and consistent performance in all three metrics, with precision, recall, and F1-scores close to 0.82, as we can see in Fig.~\ref{doc:metrics_bots}.
The introduction of controlled stochasticity into the reward function improves the robustness of the policy and
prevents overfitting to deterministic reward patterns, producing a reliable containment strategy under uncertain traffic conditions.

Overall, these results confirm that the proposed 2-layer DRL agent can be adapted to different SOC operational objectives through reward shaping alone, without modifying the underlying model
architecture to assist SOC analysts in decision making to either \texttt{Contain} or \texttt{Allow} traffic.

\subsection{Decision Cost and Regret Analysis on Boss of SOC Dataset}
\label{subsec:cost_regret}

\paragraph{Setup}
Let each aggregated time window (state) be indexed by $t \in \{1,\dots,T\}$.
The DRL policy $\pi_\theta$ outputs a containment decision
$a_t \in \{0,1\}$, where $a_t=1$ indicates \texttt{containment}(raise an alert/take action) and $a_t=0$ denotes \texttt{allow} (do nothing).
The ground-truth label is $y_t \in \{0,1\}$, where $y_t=1$ indicates malicious activity and $y_t=0$ indicates legitimate activity.
Each decision produces one of four outcomes:
True Positive (TP): $(a_t,y_t)=(1,1)$,
False Positive (FP): $(a_t,y_t)=(1,0)$,
False Negative (FN): $(a_t,y_t)=(0,1)$,
True Negative (TN): $(a_t,y_t)=(0,0)$.

\paragraph{Decision cost model:} In the SOC environment, it is important to know the cost associated with each decision.
We assign an operational cost to each outcome to reflect the use and risk of SOC resources.
Let $C_{\mathrm{TP}}, C_{\mathrm{FP}}, C_{\mathrm{FN}}, C_{\mathrm{TN}} \in \mathbb{R}$ denote the
cost (negative values may represent \emph{benefit} or \emph{cost savings}).
We define the per-window decision cost as:
\begin{equation}
\label{eq:decision_cost}
c_t \;=\;
\begin{cases}
C_{\mathrm{TP}}, & a_t=1,\; y_t=1 \quad (\mathrm{TP}),\\
C_{\mathrm{FP}}, & a_t=1,\; y_t=0 \quad (\mathrm{FP}),\\
C_{\mathrm{FN}}, & a_t=0,\; y_t=1 \quad (\mathrm{FN}),\\
C_{\mathrm{TN}}, & a_t=0,\; y_t=0 \quad (\mathrm{TN}).
\end{cases}
\end{equation}
The \texttt{total decision cost} and \texttt{average decision cost} over a fold are represented by:
\begin{equation}
\label{eq:total_avg_cost}
C_{\mathrm{total}} \;=\; \sum_{t=1}^{T} c_t,
\qquad
\overline{C} \;=\; \frac{1}{T}\sum_{t=1}^{T} c_t.
\end{equation}

\paragraph{Regret analysis}
Regret analysis evaluates how much less optimal the policy's decision is compared to an \texttt{oracle} that always selects the action with the minimal cost for that particular activity.
For each time window $t$, the oracle decision cost is represented by:
\begin{equation}
\label{eq:oracle_cost}
c_t^\star \;=\; \min_{a \in \{0,1\}} \; c(y_t,a),
\end{equation}
where $c(y,a)$ follows Eq.~\eqref{eq:decision_cost}.
The instantaneous regret is represented by:
\begin{equation}
\label{eq:instant_regret}
r_t \;=\; c_t \;-\; c_t^\star \;\;\ge\; 0,
\end{equation}
and the \emph{total} and \emph{average regret} are calculated by:
\begin{equation}
\label{eq:total_avg_regret}
R_{\mathrm{total}} \;=\; \sum_{t=1}^{T} r_t,
\qquad
\overline{R} \;=\; \frac{1}{T}\sum_{t=1}^{T} r_t.
\end{equation}

The lower regret shows that the policy is close to the oracle in terms of operational cost, i.e., it makes fewer mistakes (specifically false positives and false negatives under the chosen cost profile).

\paragraph{Decision cost and regret evaluation}
Table~\ref{tab:cost_regret_results} shows the averaged operational decision cost and regret calculated across different reward modes.
As we can see in Table~\ref{tab:cost_regret_results} Mode C and D achieve the lowest average decision cost close to $-0.79$, indicating improved operational efficiency compared to Mode A and B.
Moreover, Mode D achieves the lowest average regret of $1.358$, highlighting that stochastic reward improves the robustness of the policy during shifts in temporal distributions.
In contrast, Mode B shows the highest regret $2.588$ and variability across folds with a standard deviation of $3.059$, highlighting the sensitivity to false positive penalties. 
Overall, these results show that the balanced reward policy of Modes C and D offers stable containment policies across evolving traffic compared to strictly recall-oriented (Mode A) or false-positive intolerant (Mode B) configurations.


\begin{table}[t]
\centering
\caption{Average Decision Cost and Regret Across Time-Series Folds}
\label{tab:cost_regret_results}
\begin{tabular}{lcccc}
\hline
Mode & Mean Decision Cost & Std. Dev. & Mean Regret & Std. Dev. \\
\hline
A & -0.526 & 1.049 & 1.474 & 1.049 \\
B & -0.254 & 1.126 & 2.588 & 3.059 \\
C & -0.789 & 1.233 & 1.684 & 1.921 \\
D & -0.794 & 1.027 & 1.358 & 1.250 \\
\hline
\end{tabular}
\end{table}


\subsection{Percentage Reduction in Traffic Flows Forwarded to LLM}

The percentage reduction in traffic flows indicates how each DRL reward mode filters traffic before forwarding it to LLM for analysis.
A higher percentage of reduction signifies that the mode suppresses more traffic and reduces LLM processing overhead.
A lower percentage of reduction indicates that the mode forwards more traffic, preserving broader coverage for LLM analysis.

\begin{figure}[h]
    \includegraphics[width=\textwidth]{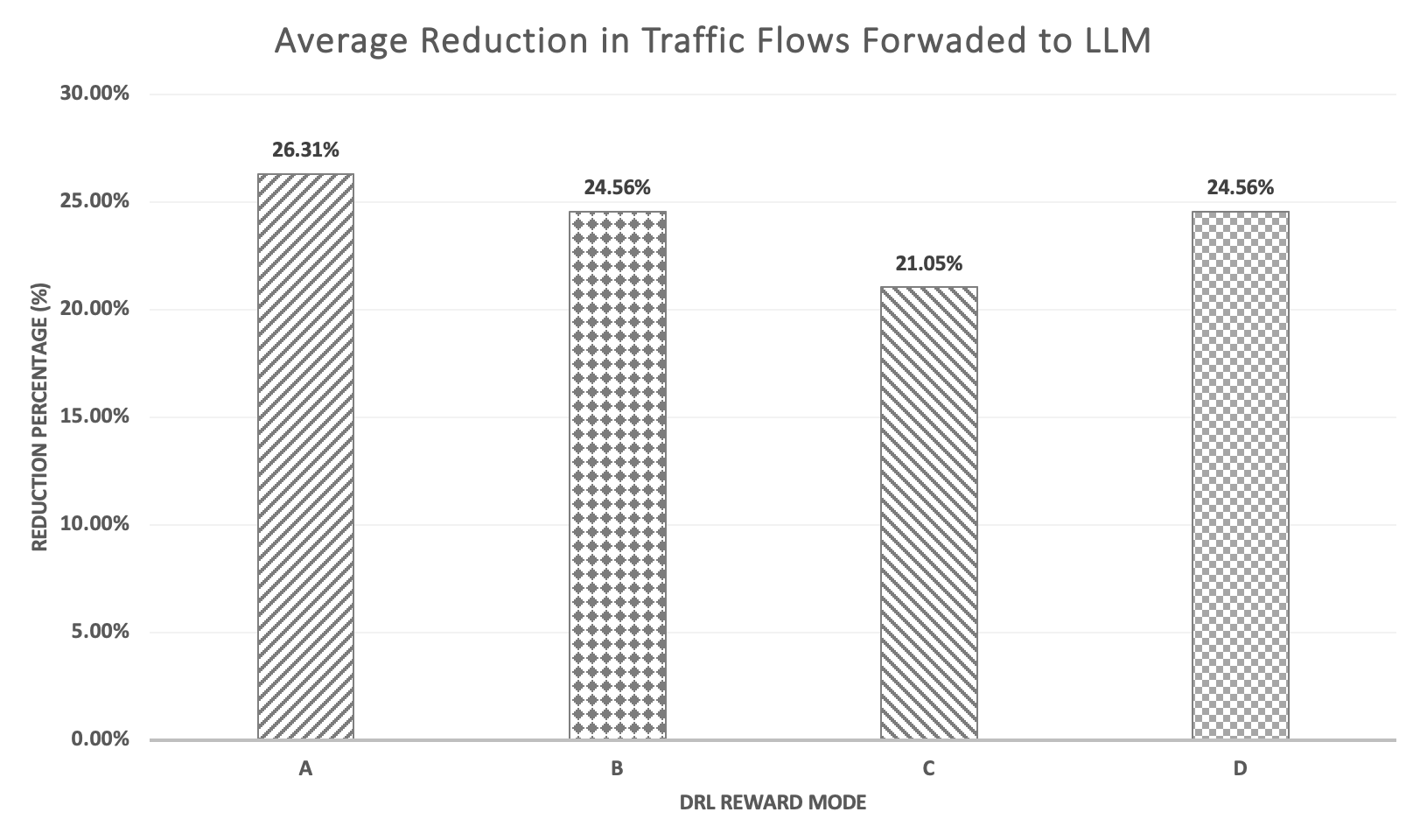}
    \centering
    \caption{Average Reduction in Traffic Flows Forwarded to LLM Across Modes (A-D)}
    \label{doc:percentage_reduction}
\end{figure}

As we can see in Fig.~\ref{doc:percentage_reduction}, \texttt{Mode A} offers more reduction in traffic flows forwarded to LLm for analysis. This is because it provides a higher penalty to false negatives than to false positives, which makes the agent sensitive to malicious activity, while still allowing some filtering of traffic. 
\texttt{Mode B} penalizes false positives and rewards true negatives both strongly. 
This encourages the DRL agent to suppress benign traffic, which supports traffic reduction and avoids unnecessary downstream analysis.
\texttt{Mode C} is a more balanced reward structure between malicious detection and benign traffic suppression.
As it is less filtering than the more selective modes, it tends to keep more traffic for downstream LLM analysis, leading to a lower percentage reduction.
\texttt{Mode D} is less conservative and has some exploratory reward design. 
So, it is less rigid in its traffic suppression, producing a moderate level of alert reduction while still allowing broader investigation of suspicious traffic flows.

\subsection{Policy Adaptation Across Reward Modes with Simulated Dataset}

\begin{figure}[h]
    \includegraphics[width=\textwidth]{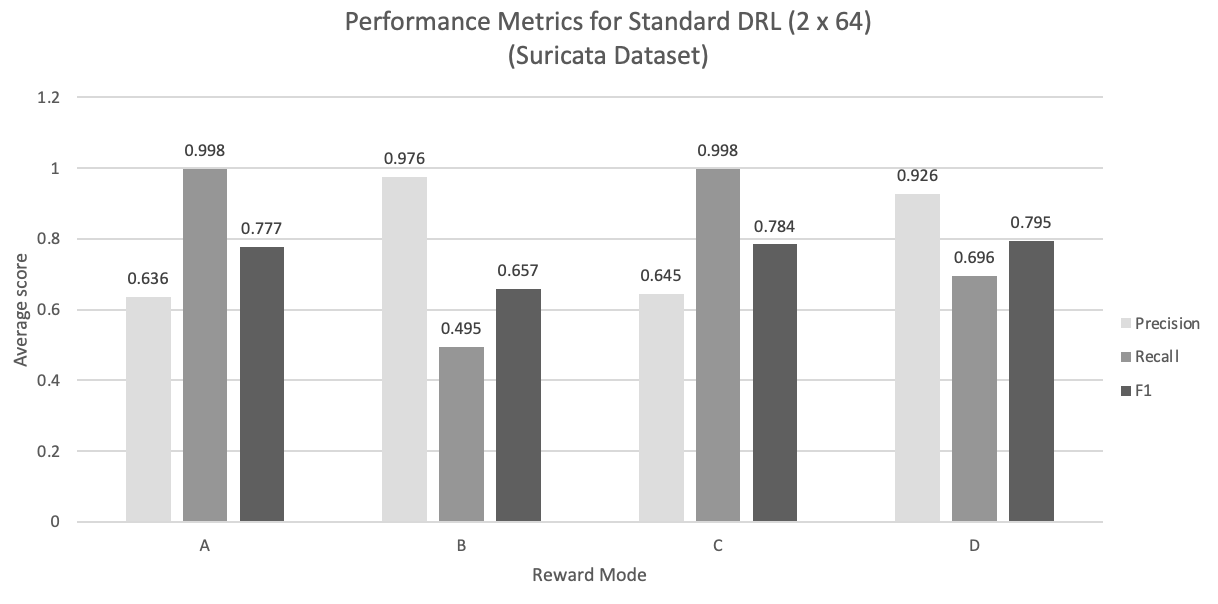}
    \centering
    \caption{Performance Evaluation of Across Modes (A-D) on Simulated Dataset}
    \label{doc:drl_suricata_metrics}
\end{figure}

We also simulated the threat model mentioned in Section~\ref{sec:threat_model}.
Fig.~\ref{doc:drl_suricata_metrics} shows the performance of the DRL agent with a two-layer architecture (2$\times$64) in the simulated Suricata traffic in the four reward modes. 
As we can see in Fig.~\ref{doc:drl_suricata_metrics}, it clearly demonstrates that the reward function impacts the agent’s
containment behavior under controlled traffic conditions.

As shown in Fig.~\ref{doc:drl_suricata_metrics}, mode A achieves a recall of $0.998$, resulting in almost perfect detection of malicious activity.
However, the precision is $0.636$ in mode A. 
It indicates that the agent strongly flags suspicious traffic, which is desirable in early detection scenarios, but causes alert fatigue to SOC analysts due to an increased number of false positives. 
Mode B achieves high precision $0.976$ while significantly reducing recall to $0.495$. 
It validates that the heavy penalization of false positives in Mode B causes a conservative containment policy that prioritizes high-confidence alerts and allows a substantial part of malicious flows to pass.
It reduces the alert fatigue on SOC analysts due to the high number of false positives.

Mode C achieves a more balanced performance, with a recall of $0.998$ and a precision of $0.645$, resulting in an F1-score of 0.784. 
It indicates that Mode C captures malicious activity while maintaining some control over false positives, making it appropriate for environments that require high detection and operational stability. 
Finally, Mode D shows consistent and well-balanced behavior, with
precision, recall, and F1-scores of $0.926$, $0.696$, and $0.795$ respectively.
It reflects that the stochasticity in the reward function enhances robustness by preventing the agent from overfitting to deterministic traffic patterns commonly present in simulated datasets.

Overall, these results point that the proposed DRL agent adapts its containment strategy in response to different reward modes, validating the effectiveness of reward shaping to model different SOC objectives.

\subsection{Percentage Reduction in Simulated Traffic Forwarded to LLM}

\begin{figure}[h]
    \includegraphics[width=\textwidth]{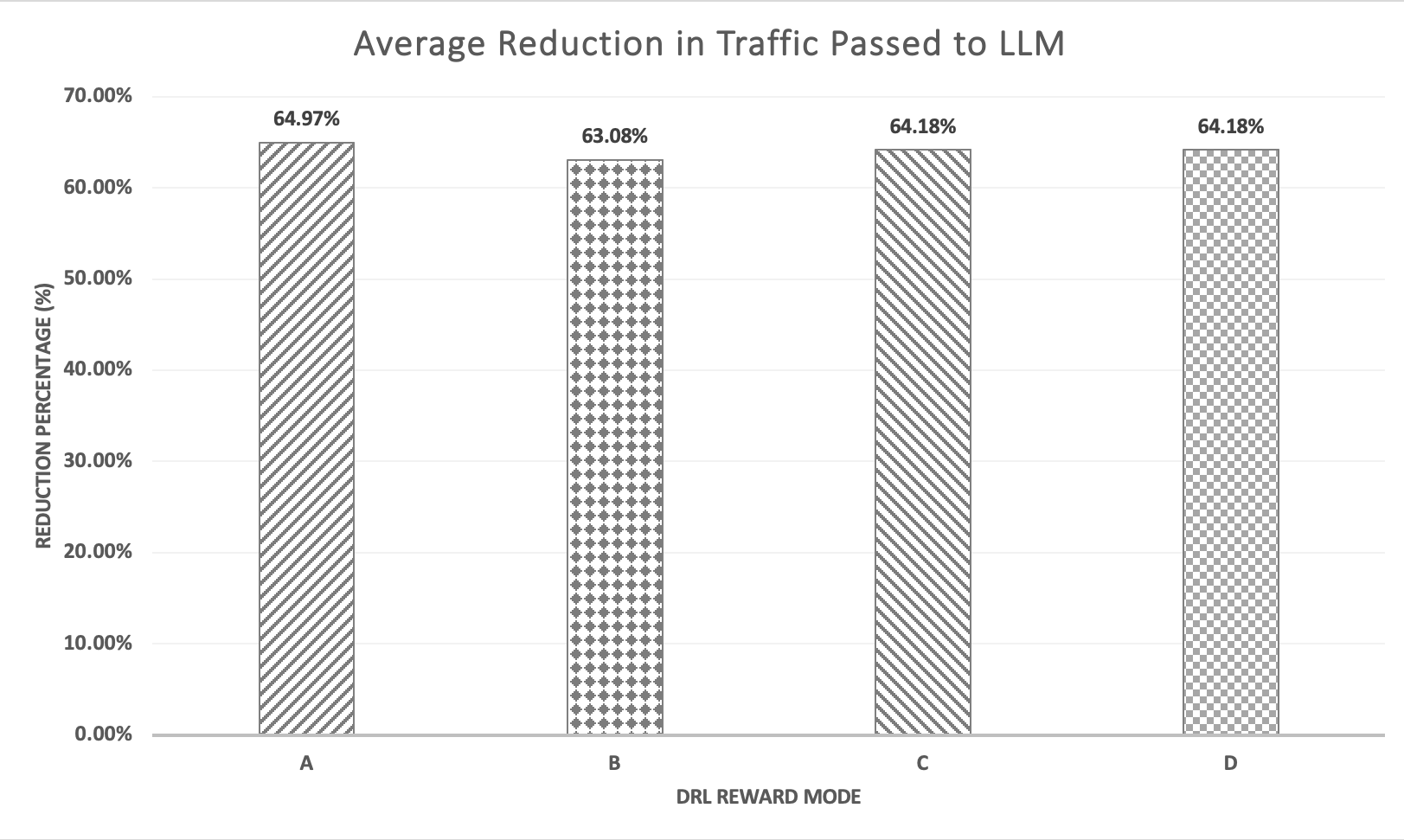}
    \centering
    \caption{Average Reduction in Traffic Forwarded to LLM for Analysis}
    \label{doc:average_reduction_suricata}
\end{figure}

In the framework, alert capture rate is equivalent to recall, as both evaluate the proportion of true alert instances that are successfully forwarded to the LLM for processing.
As we can see in Fig.~\ref{doc:average_reduction_suricata}, across all reward modes, the DRL triage mechanism reduces traffic forwarded to the LLM by approximately 63\%-65\%.
However, this reduction is associated with different recall outcomes, as shown in Fig~\ref{doc:drl_suricata_metrics}.
As shown in Fig.~\ref{doc:average_reduction_suricata} and Fig.~\ref{doc:drl_suricata_metrics}, Modes A and C forward almost all true traffic alerts while still achieving strong traffic reduction, indicating an effective balance between efficiency and detection performance.
Mode B also obtains a similar reduction in traffic forwarded to LLM processing but with lower recall, highlighting that its policy is filtering too much. 
Mode D performs moderately, but remains less effective than Modes A and C in maintaining traffic flow coverage.
This design ensures that the LLM is not overloaded with a huge traffic volume for analysis.

\subsection{SOC Analyst Validation via Splunk Analysis on Simulated Dataset}

\begin{figure}[h]
    \includegraphics[width=\textwidth]{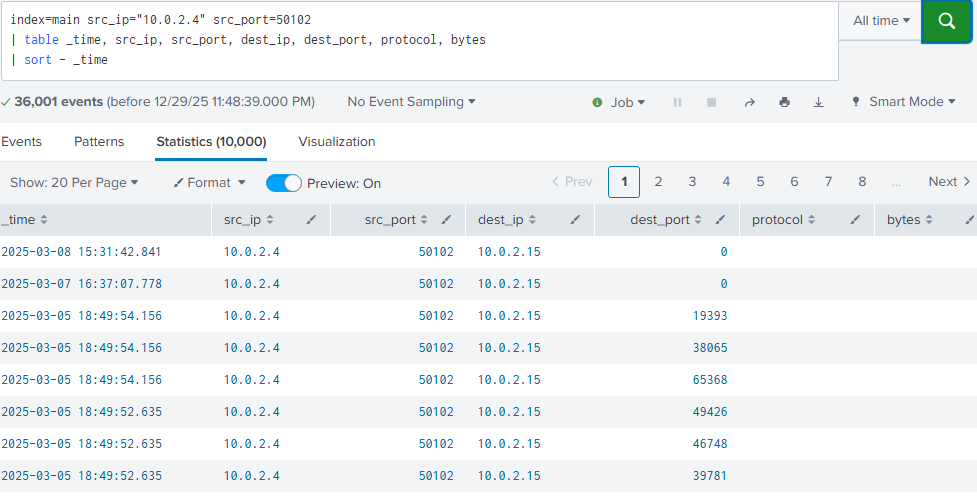}
    \centering
    \caption{Analysis of flows from malicious host}
    \label{doc:splunk_malicious_flows}
\end{figure}

In the framework, after suspicious traffic is prioritized using the DRL network and AAD score, an LLM-based multi-agent system provides contextual insights and generates SPL queries to validate the traffic on Splunk.
As mentioned in Section~\ref{sec:methodology}, the Senior SOC Triage Analyst agent provides contextual insights and generate SPL queries for the analysis. 
Fig.~\ref{doc:splunk_malicious_flows} shows the SPL query applied to filter malicious traffic on Splunk.
The SPL query filters the traffic originating from the flagged host machine with the IP address of $10.0.2.4$, allowing the SOC analyst to analyze the communication pattern.
For clear representation, we have shown the IP addresses in the paper, but LLM triage is performed on anonymized data.
This design ensures that the LLM reasons over traffic behavior and context
rather than sensitive identifiers, enabling privacy-preserving SOC triage
without degrading analytical capability.

As we can see in Fig.~\ref{doc:splunk_malicious_flows}, repetitive communication from the same source IP address (10.0.2.4) towards the destination IP address (10.0.2.15) on different destination ports in a very short time window highlights network scanning activity. 
It is mapped to the MITRE ATT\&CK technique of T1046.
The filtered traffic on Splunk does not show protocol violation or malformed packets, rather the anomaly is identified through temporal aggregation and behavioral pattern, signifying the advantage of DRL based containment decision over signature driven detection.

It is important to note that during LLM assisted triage the majority of prioritized flows originate from the same host, therefore, a detailed table highlighting LLM agent determination along with MITRE ID is omitted from the paper for brevity.
This is because in the simulation we used Kali Linux with the same IP address.
From a threat hunting perspective, validating the DRL-prioritized traffic through independent SIEM-level behavioral pattern demonstrates that the proposed framework not only achieves high detection performance, but also produces explainable insights and assists SOC analyst in decision making to either allow the traffic in the network, block, or monitor it for a while. 
The whole process aligns with the real world SOC workflows and ensures that SOC analyst verifies DRL decision and LLM insights on the SIEM tool such as Splunk rather than only relying on the LLM.


\section{Discussion}
\label{sec:discussion}

Our framework provides a multi-layer threat detection architecture that combines Deep Reinforcement Learning (DRL), Autoencoder based anomaly detection (AAD), and LLM based agents for triage and automated SOC decision along with verification of the results using SIEM tool such as \texttt{Splunk}.
In the framework, once the DRL agent is trained on aggregated traffic features and learn containment decision, then AAD score is used along with the DRL decision in prioritizing flows for LLM triage.
It reduces the processing overhead on LLM by only forwarding the prioritized flows for analysis that also prevents hallucination.  
The LLM agents also generate SPL queries that are used for filtering malicious flows on the Splunk dashboard.
It is helpful for junior SOC analysts who are not well versed in writing SPL queries.
Moreover, in the agentic AI framework, another LLM agent generates an incidence response playbook and provides mapping with the MITRE ATT\&CK ID.
In contrast to manual investigation on the SIEM tool, such as Splunk, our agentic AI framework automates detection, prioritization, contextual analysis, and report generation for detailed investigation.

The proposed framework is aligned with real-world SOC workflows.
The DRL agent learns sequential decision policies that directly model containment actions under uncertainty rather than handling detection as a static classification problem. 
This formulation shows operational SOC decision-making, where actions carry costs and delayed consequences. 
The use of multiple reward profiles further demonstrates the flexibility of the framework in adapting to different organizational risk tolerances, such as prioritizing low false-positive rates in high-volume environments.
The current experimental evaluation focuses on binary containment decisions that \texttt{containment} and \texttt{no action (allow)}.
However, SOC environment often contains actions in multiple stages such as: monitor, throttle, escalate, and isolate. 
Extending the action would cover more practical scenarios and will also increase the complexity in the reward function.

One may argue that in the framework, the use of public LLM for triage requires providing sensitive security data to an external provider.
However, local LLM can be used to avoid forwarding sensitive logs to public LLM for triage. 
Another solution is to use data anonymization technique to anonymize sensitive information such as IP addresses before sending it to the public LLM for analysis.

Another important concern of the current framework is scalability, as it is common for tens of thousands of traffic logs to arrive per second in security operation centers.
Our framework addresses this challenge through triage priority after the DRL decision, ensuring that the agentic analysis is only applied to the high priority subset of traffic flows.
Raw network telemetry is aggregated into a fixed temporal time window, and flows are summarized using statistical descriptors such as the mean \& maximum value of ports, in bytes, out bytes, and protocol. 
Aggregation of flows reduces the dimensionality of the data, transforming raw traffic into manageable volumes. 
The DRL agent operates as a policy level filter evaluating each aggregated window rather than individual flows.
The autoencoder anomaly detection (AAD) score is combined with the DRL decision to calculate the triage priority, which ranks only traffic windows flagged by the DRL agent for containment.
Finally, the LLM based agents are invoked only for small number of high priority traffic flows, this selective strategy makes the framework computationally feasible for SOC environment with high traffic volumes.
However, the aggregated time window feature may limit the detection of short lived attacks.
The experimental evaluation demonstrate that combining Deep Reinforcement Learning and autoencoder anomaly detection along with contextual analysis improves the overall efficiency and decision making of SOC analysts.

Moreover, currently the proposed framework lacks complete automation from detection to mitigation. 
The main purpose is to assist the SOC analyst in identifying the suspicious traffic for investigation and help them in decision making.
The mitigation of malicious traffic is performed by SOC analysts after investigating on the SIEM tool.
Moreover, the aim is also to study how agentic AI coupled with the SIEM tool can assist SOC analysts in threat hunting.
Rather than complete automation from monitoring to mitigation at this stage, it is nice to have automation across key security workflows, for improved performance.

\section{Conclusion and Future Work}
\label{sec:conclusion_future_work}

This paper presented a threat hunting framework based on Agentic AI that integrates Deep Reinforcement Learning (DRL), Autoencoder-based Anomaly Detection (AAD), and LLM driven multi-agent contextual analysis.
The framework represents the SOC environment as a sequential decision-making problem, where actions are decided based on learned policies instead of being treated as static classification outputs. 
To demonstrate the feasibility and effectiveness of our proposed framework, we developed a proof-of-concept prototype and evaluated both on the public dataset as well as simulated dataset.
The results show that reward shaping allows the DRL agent to adapt its containment decision to different SOC objectives, such as maximizing recall, minimizing false positives, and balancing detection.
The operational cost and regret analysis are also incorporated into the learning process that provides a more realistic evaluation of the SOC environment. 

Our experiment and analysis of the prototype have identified some key benefits of the framework, (1) The modular architecture of the framework separates the responsibilities across different components; (2) The DRL agent works on aggregated raw traffic features and is trained without AAD score, preventing feature leakage; (3) The AAD score is applied only after the DRL decision to prioritize flagged windows for LLM triage to avoid processing overhead on LLM; (4) The integration of DRL–based decision with anomaly-aware prioritization and LLM-driven reasoning offers explainable, analyst-aligned threat hunting outcomes; (5) The reduced traffic volume forwarded to the LLM improves causes less processing overhead on the LLM.
This multi-layer architecture enables scalability by ensuring that computationally expensive LLM reasoning is applied only where it is most needed.
As the SOC environment operates with multi-stage actions such as monitoring, escalating, isolating traffic, etc.
In our future work, we will extend the actions space to support multi-level containment decisions.
Currently, the framework relies on fixed time-window aggregation due to which it can be difficult to detect short-lived malicious traffic. 
In our future work, we will explore adaptive window strategies and hierarchical policies that work at multiple temporal level.
Moreover, we will also investigate the use of domain specific LLM or locally deployed LLM to improve reliability, reduce hallucination risk and address data privacy concerns.
Furthermore, we will investigate ethical considerations, including transparency, bias, and accountability in agentic SOC systems, to support responsible deployment of autonomous decision-making systems for cybersecurity operations.
Finally, we will perform detailed experimentation with multiple heterogeneous log sources and evaluate the performance.

\bibliographystyle{elsarticle-num}
\bibliography{references}

\appendix
\section{Mathematical Details and Numerical Illustration}
\label{app:math_details}

This appendix provides the detailed mathematical formulation and a worked numerical example
supporting the reinforcement learning–based containment framework described in Section~5.
These details are included for completeness and reproducibility and are not required for
understanding the main results.

\subsection{Time-Windowed State Construction}
\label{app:state_construction}

Let raw network traffic be represented as a sequence of flow records:
\begin{equation}
\mathcal{F} = \{ f_1, f_2, \dots, f_N \}.
\end{equation}

Traffic is aggregated into non-overlapping windows of duration $\Delta t$ (5 minutes).
For each window $w_t$, a state vector $s_t$ is constructed using statistical aggregation.

For numerical attributes
$x \in \{\text{src\_port}, \text{dest\_port}, \text{bytes\_in}, \text{bytes\_out}\}$,
we compute:
\begin{equation}
\mu(x)_t = \frac{1}{|w_t|} \sum_{f_i \in w_t} x_i,
\end{equation}
\begin{equation}
\max(x)_t = \max_{f_i \in w_t} x_i.
\end{equation}

The mean captures typical behavior within the window, while the maximum captures bursty or
extreme activity commonly associated with scanning, flooding, or data exfiltration.

Categorical features (source IP, destination IP, protocol) are encoded using reduced-cardinality
one-hot representations and aggregated as occurrence counts per window.

The resulting state vector is:
\begin{equation}
s_t =
\big[
\mu(\text{src\_port}),
\max(\text{src\_port}),
\mu(\text{dest\_port}),
\max(\text{dest\_port}),
\mu(\text{bytes\_in}),
\max(\text{bytes\_out}),
\mathbf{IP}_{src},
\mathbf{IP}_{dst},
\mathbf{Proto}
\big]_t.
\end{equation}


\subsection{Worked Numerical Example}
\label{app:numerical_example}

Consider a single 5-minute window with the following aggregated numerical features:
\begin{align*}
\mu(\text{src\_port}) &= 443, \\
\max(\text{src\_port}) &= 443, \\
\mu(\text{dest\_port}) &= 51{,}024, \\
\max(\text{dest\_port}) &= 51{,}083, \\
\mu(\text{bytes\_in}) &= 1{,}420, \\
\max(\text{bytes\_out}) &= 98{,}230.
\end{align*}

After concatenation with encoded categorical features, the state vector is $s_t \in \mathbb{R}^d$.

Assume the policy network produces logits:
\begin{equation}
z = [-1.31, \; 1.11],
\end{equation}
corresponding to \emph{no containment} and \emph{containment}, respectively.

Applying softmax:
\begin{align*}
\pi(0 \mid s_t) &= \frac{e^{-1.31}}{e^{-1.31}+e^{1.11}} \approx 0.08,\\
\pi(1 \mid s_t) &= \frac{e^{1.11}}{e^{-1.31}+e^{1.11}} \approx 0.92.
\end{align*}

The agent selects $a_t = 1$ (containment) with confidence $0.92$.

If the ground-truth label is malicious ($y_t = 1$), the outcome is a true positive and the
corresponding reward is assigned according to the active reward profile.

\subsection{Decision Cost and Regret}
\label{app:cost_regret}

To complement standard detection metrics, we quantify operational impact using decision cost
and regret.

The decision cost at time $t$ is defined as the negative of the obtained reward:
\begin{equation}
\mathrm{cost}_t = -r_t.
\end{equation}

This formulation directly reflects operational burden: high cost corresponds to undesirable
actions such as false positives or missed detections.

Regret measures how far the agent’s decision deviates from the best possible action under the
defined reward function. We define regret as:
\begin{equation}
\mathrm{regret}_t = r_t^\star - r_t,
\end{equation}
where
\begin{equation}
r_t^\star = \max_{a \in \{0,1\}} \mathcal{R}(a, y_t)
\end{equation}
is the oracle reward assuming perfect knowledge of the ground-truth label.

A regret value of zero indicates an optimal decision, while larger values indicate increasingly
costly deviations from the ideal SOC response.


\subsection{Worked Numeric Example (End-to-End): Aggregation $\rightarrow$ AAD Score $\rightarrow$ DRL Action Probability}

\noindent\textbf{Goal.} This example shows (i) how raw flow logs inside a 5-minute window are converted into fixed-length features (mean/max), (ii) how an Autoencoder-based Anomaly Detection (AAD) score is computed from reconstruction error, and (iii) how a 2-layer DRL policy converts the feature vector into a containment probability (e.g., $0.92$).

\subsubsection{Step 1: Raw flows in a 5-minute window}
Assume within a 5-minute time window $w$, we observe $N=4$ flow records with numeric attributes:
\[
\text{src\_port},\ \text{dest\_port},\ \text{bytes\_in},\ \text{bytes\_out}.
\]
Let the four flows be:
\[
\begin{array}{c|cccc}
\hline
\text{Flow } i & \text{src\_port}_i & \text{dest\_port}_i & \text{bytes\_in}_i & \text{bytes\_out}_i\\
\hline
1 & 443 & 52344 & 1200 & 300\\
2 & 443 & 52345 & 1300 & 350\\
3 & 22  & 52346 & 8000 & 9000\\
4 & 22  & 52347 & 9000 & 11000\\
\hline
\end{array}
\]

\subsubsection{Step 2: Window aggregation (mean and max)}
We compute a fixed-length feature vector $\mathbf{x}_w \in \mathbb{R}^{6}$ using mean and max statistics:
\[
\mathbf{x}_w =
\big[
\text{src\_port\_mean},\ \text{src\_port\_max},\ 
\text{dest\_port\_mean},\ \text{dest\_port\_max},\
\text{bytes\_in\_mean},\ \text{bytes\_out\_max}
\big].
\]
The aggregation operators are:
\[
\text{mean}(v)=\frac{1}{N}\sum_{i=1}^{N} v_i,\qquad
\text{max}(v)=\max_{i\in\{1,\dots,N\}} v_i.
\]
Compute each component:

\begin{itemize}
    \item[ ] src\_port\_mean = $\frac{443 + 443 + 22 + 22}{4}$ = 232.5
    \item[ ] src\_port\_max = $\max(443,443,22,22)$ = 443
    \item[ ] dest\_port\_mean = $\frac{52344+52345+52346+52347}{4}$ = 52345.5,
    \item[ ] dest\_port\_max = $\max(52344,52345,52346,52347)$ = 52347
    \item[ ] bytes\_in\_mean = $\frac{1200+1300+8000+9000}{4}$ = 4875
\item[ ] bytes\_out\_max = $\max(300,350,9000,11000)$ = 11000
\end{itemize}




Thus,
\[
\mathbf{x}_w = [232.5,\ 443,\ 52345.5,\ 52347,\ 4875,\ 11000].
\]

\subsubsection{Step 3: Standardization (as used in training)}
Before AAD/DRL, numeric features are standardized:
\[
\tilde{\mathbf{x}}_w = \frac{\mathbf{x}_w - \boldsymbol{\mu}}{\boldsymbol{\sigma}},
\]
where $\boldsymbol{\mu}$ and $\boldsymbol{\sigma}$ are computed from the \emph{training} data only.
For illustration, assume:
\[
\boldsymbol{\mu}=[200,\ 400,\ 52000,\ 52000,\ 2000,\ 5000],
\quad
\boldsymbol{\sigma}=[100,\ 100,\ 200,\ 200,\ 2000,\ 3000].
\]
Then:
\[
\tilde{\mathbf{x}}_w=
\left[
\frac{232.5-200}{100},\
\frac{443-400}{100},\
\frac{52345.5-52000}{200},\
\frac{52347-52000}{200},\
\frac{4875-2000}{2000},\
\frac{11000-5000}{3000}
\right].
\]
Numerically:
\[
\tilde{\mathbf{x}}_w=
[0.325,\ 0.43,\ 1.7275,\ 1.735,\ 1.4375,\ 2.0].
\]

\subsubsection{Step 4: AAD score via Autoencoder reconstruction error}
\paragraph{AAD model.}
Let the AAD be a autoencoder trained on early benign data:
\[
f_{\theta}:\mathbb{R}^{6}\rightarrow\mathbb{R}^{6}.
\]
We use a hidden layer encoder, bottleneck dimension $d=2$, and a decoder:
\[
\mathbf{h}=\phi(\mathbf{W}_1\tilde{\mathbf{x}}_w+\mathbf{b}_1),\quad
\mathbf{z}=\phi(\mathbf{W}_2\mathbf{h}+\mathbf{b}_2),
\]
\[
\hat{\mathbf{h}}=\phi(\mathbf{W}_3\mathbf{z}+\mathbf{b}_3),\quad
\hat{\mathbf{x}}_w=\mathbf{W}_4\hat{\mathbf{h}}+\mathbf{b}_4,
\]
where $\phi(\cdot)=\max(0,\cdot)$ is ReLU, and $\hat{\mathbf{x}}_w$ is the reconstruction.

\paragraph{Concrete numeric forward pass (Example)}
For the example, we illustrate with a small hidden size (the real implementation can use larger widths).
Assume:
\[
\mathbf{W}_1=
\begin{bmatrix}
0.30&0.10&0.05&0.05&0.10&0.20\\
0.10&0.20&0.10&0.10&0.20&0.10\\
0.05&0.05&0.30&0.30&0.10&0.10\\
0.10&0.10&0.20&0.20&0.10&0.05
\end{bmatrix},
\quad
\mathbf{b}_1=\mathbf{0}.
\]
Compute $\mathbf{h} = \phi(\mathbf{W}_1\tilde{\mathbf{x}}_w)$:
\[
\mathbf{W}_1\tilde{\mathbf{x}}_w=
\begin{bmatrix}
0.30(0.325)+0.10(0.43)+0.05(1.7275)+0.05(1.735)+0.10(1.4375)+0.20(2.0)\\
0.10(0.325)+0.20(0.43)+0.10(1.7275)+0.10(1.735)+0.20(1.4375)+0.10(2.0)\\
0.05(0.325)+0.05(0.43)+0.30(1.7275)+0.30(1.735)+0.10(1.4375)+0.10(2.0)\\
0.10(0.325)+0.10(0.43)+0.20(1.7275)+0.20(1.735)+0.10(1.4375)+0.05(2.0)
\end{bmatrix}.
\]
Numerically:
\[
\mathbf{W}_1\tilde{\mathbf{x}}_w \approx
\begin{bmatrix}
0.0975+0.043+0.0864+0.0868+0.1438+0.4000\\
0.0325+0.0860+0.1728+0.1735+0.2875+0.2000\\
0.0163+0.0215+0.5183+0.5205+0.1438+0.2000\\
0.0325+0.0430+0.3455+0.3470+0.1438+0.1000
\end{bmatrix}
=
\begin{bmatrix}
0.8575\\
0.9523\\
1.4204\\
1.0118
\end{bmatrix}.
\]
After ReLU:
\[
\mathbf{h}=[0.8575,\ 0.9523,\ 1.4204,\ 1.0118]^{\top}.
\]

Now, we define bottleneck weights as:
\[
\mathbf{W}_2=
\begin{bmatrix}
0.6&0.2&0.1&0.1\\
0.1&0.2&0.6&0.1
\end{bmatrix},
\quad
\mathbf{b}_2=\mathbf{0}.
\]
Compute the bottleneck:
\[
\mathbf{z}=\phi(\mathbf{W}_2\mathbf{h})=
\phi\left(
\begin{bmatrix}
0.6(0.8575)+0.2(0.9523)+0.1(1.4204)+0.1(1.0118)\\
0.1(0.8575)+0.2(0.9523)+0.6(1.4204)+0.1(1.0118)
\end{bmatrix}
\right).
\]
Numerically:
\[
\mathbf{z}\approx
\phi\left(
\begin{bmatrix}
0.5145+0.1905+0.1420+0.1012\\
0.0858+0.1905+0.8522+0.1012
\end{bmatrix}
\right)
=
\begin{bmatrix}
0.9482\\
1.2297
\end{bmatrix}.
\]

For reconstruction, we assume a simple linear decoder (Example):
\[
\hat{\mathbf{x}}_w =
\mathbf{A}\mathbf{z},
\quad
\mathbf{A}=
\begin{bmatrix}
0.20&0.00\\
0.10&0.05\\
0.30&0.10\\
0.30&0.10\\
0.10&0.20\\
0.05&0.40
\end{bmatrix}.
\]
Then:
\[
\hat{\mathbf{x}}_w=
\begin{bmatrix}
0.20(0.9482)+0.00(1.2297)\\
0.10(0.9482)+0.05(1.2297)\\
0.30(0.9482)+0.10(1.2297)\\
0.30(0.9482)+0.10(1.2297)\\
0.10(0.9482)+0.20(1.2297)\\
0.05(0.9482)+0.40(1.2297)
\end{bmatrix}
\approx
\begin{bmatrix}
0.1896\\
0.1563\\
0.4074\\
0.4074\\
0.3408\\
0.5393
\end{bmatrix}.
\]

\paragraph{AAD score as mean squared reconstruction error}

Define AAD as:
\[
\text{AAD}(\tilde{\mathbf{x}}_w)=
\frac{1}{6}\sum_{j=1}^{6}\left(\tilde{x}_{w,j}-\hat{x}_{w,j}\right)^2.
\]
Compute the per-dimension errors as:
\[
\tilde{\mathbf{x}}_w-\hat{\mathbf{x}}_w
=
[0.325-0.1896,\ 0.43-0.1563,\ 1.7275-0.4074,\ 1.735-0.4074,\ 1.4375-0.3408,\ 2.0-0.5393].
\]
Numerically:
\[
\tilde{\mathbf{x}}_w-\hat{\mathbf{x}}_w \approx
[0.1354,\ 0.2737,\ 1.3201,\ 1.3276,\ 1.0967,\ 1.4607].
\]
Squared errors:
\[
[0.0183,\ 0.0749,\ 1.7427,\ 1.7625,\ 1.2028,\ 2.1336].
\]
Finally:
\[
\text{AAD}(\tilde{\mathbf{x}}_w)\approx
\frac{0.0183+0.0749+1.7427+1.7625+1.2028+2.1336}{6}
=
\frac{6.9348}{6}
\approx 1.1558.
\]
\noindent\textbf{Interpretation:} Higher AAD shows a larger deviation from benign reconstruction, i.e., a more anomalous traffic window.

\subsubsection{Step 5: DRL policy converts features into a containment probability}
Let the DRL policy network (2 hidden layers; in the implementation 64 neurons per layer) output two logits for actions:
\[
a\in\{0,1\}\quad\text{where }a=1\ \text{means containment}.
\]
We denote:
\[
\mathbf{h}^{(1)}=\phi(\mathbf{W}^{(1)}\tilde{\mathbf{x}}_w+\mathbf{b}^{(1)}),\qquad
\mathbf{h}^{(2)}=\phi(\mathbf{W}^{(2)}\mathbf{h}^{(1)}+\mathbf{b}^{(2)}),
\]
\[
\boldsymbol{\ell}=\mathbf{W}^{(3)}\mathbf{h}^{(2)}+\mathbf{b}^{(3)}
=
\begin{bmatrix}
\ell_0\\
\ell_1
\end{bmatrix},
\]
where $\ell_0$ is the logit for \emph{no containment} and $\ell_1$ is the logit for \emph{containment}.

\paragraph{Concrete numeric logits (Example)}
Let's assume the network produces logits as:
\[
\ell_0=-1.2,\qquad \ell_1=1.6.
\]
These are not probabilities; they are raw preference scores.

\paragraph{Softmax to obtain action probabilities.}
The policy probability is give by:
\[
\pi(a=i\mid \tilde{\mathbf{x}}_w)=
\frac{\exp(\ell_i)}{\exp(\ell_0)+\exp(\ell_1)},\quad i\in\{0,1\}.
\]
Compute:
\[
\exp(\ell_0)=e^{-1.2}\approx 0.301,\qquad
\exp(\ell_1)=e^{1.6}\approx 4.953.
\]
Sum:
\[
Z=\exp(\ell_0)+\exp(\ell_1)\approx 0.301+4.953=5.254.
\]
Thus:
\[
\pi(a=1\mid \tilde{\mathbf{x}}_w)=\frac{4.953}{5.254}\approx 0.943,
\qquad
\pi(a=0\mid \tilde{\mathbf{x}}_w)=\frac{0.301}{5.254}\approx 0.057.
\]
\noindent\textbf{Interpretation:} The DRL policy assigns $\approx 94.3\%$ confidence to containment for this window.

\subsubsection{Step 6: Combining DRL decision with AAD for SOC triage (post-training)}
Because AAD is excluded from DRL training (to prevent feature leakage), it is used after DRL for prioritization before forwarding to LLM:
\[
\text{Priority}_w = \mathbb{I}[a_w=1]\cdot \text{AAD}(\tilde{\mathbf{x}}_w),
\]
where $\mathbb{I}[\cdot]$ is an indicator (1 if containment is selected, else 0).
If the agent selects $a_w=1$, then:
\[
\text{Priority}_w = 1 \cdot 1.1558 = 1.1558.
\]
If $a_w=0$, then:
\[
\text{Priority}_w = 0.
\]
\noindent\textbf{Interpretation:} Only traffic windows flagged by DRL are escalated, and AAD provides a score for prioritization before LLM-based contextual analysis.

\subsubsection{Variable Dictionary}
\begin{itemize}
\item $w$: a 5-minute time window.
\item $N$: number of flow records inside window $w$.
\item $\mathbf{x}_w\in\mathbb{R}^{6}$: aggregated window feature vector (mean/max).
\item $\tilde{\mathbf{x}}_w$: standardized features, using training-set mean $\boldsymbol{\mu}$ and std $\boldsymbol{\sigma}$.
\item $f_{\theta}(\cdot)$: AAD autoencoder reconstruction function.
\item $\hat{\mathbf{x}}_w$: reconstruction of $\tilde{\mathbf{x}}_w$ produced by the autoencoder.
\item $\text{AAD}(\tilde{\mathbf{x}}_w)$: anomaly score computed as mean squared reconstruction error.
\item $a\in\{0,1\}$: DRL action (0 = no containment, 1 = containment).
\item $\boldsymbol{\ell}=[\ell_0,\ell_1]^{\top}$: logits (raw scores) output by the policy network.
\item $\pi(a\mid \tilde{\mathbf{x}}_w)$: softmax policy giving the probability of each action.
\item $\text{Priority}_w$: post-training SOC triage priority (used to rank events for LLM investigation).
\end{itemize}

\end{document}